\pdfoutput=1
\RequirePackage{ifpdf}
\ifpdf % We are running pdfTeX in pdf mode
\documentclass[pdftex]{sigma}
\else
\documentclass{sigma}
\fi

\usepackage[matrix,arrow,frame,import,curve,color]{xy}

\def\C{{\mathbb C}}
\def\P{{\mathbb P}}
\def\R{{\mathbb R}}
\def\Z{{\mathbb Z}}

\def\bM{{\boldsymbol{M}}}
\def\bN{{\boldsymbol{N}}}

\def\cE{{\cal E}}
\def\cF{{\cal F}}
\def\cG{{\cal G}}
\def\cM{{\cal M}}
\def\cO{{\cal O}}

\def\vphi{{{\varphi}}}

\def\Wt{{{\widetilde{W}}}}

\newcommand\gammah{\widehat{\gamma}}
\newcommand\deltah{\widehat{\delta}}
\newcommand\kappah{\widehat{\kappa}}
\newcommand\sigmah{\widehat{\sigma}}

\newcommand\ah{\widehat{a}}
\newcommand\Qh{\widehat{Q}}

\def\Aut{\operatorname{Aut}}
\def\End{\operatorname{End}}
\def\im{\operatorname{im}}
\def\relint{\operatorname{relint}}
\def\Sym{\operatorname{Sym}}

\let\a=\alpha
\let\b=\beta

\newcommand{\del}{\partial}

\newcommand{\bma}{\begin{pmatrix}}
\newcommand{\ema}{\end{pmatrix}}

\newcommand{\hlf}{\frac{1}{2}}

\newcommand{\F}{{\mathbb F}}

\newcommand{\La}{\Lambda}

\newcommand{\G}{\Gamma}

\newcommand{\dd}{\delta}

\newcommand{\f}{\psi}

\def\G{\Gamma}

\def\cD{{\cal D}}

\renewcommand{\b}{\beta}
\newcommand{\g}{\gamma}

\def\d{{\rm d}}

\newcommand{\ibar}{{\bar \imath}}

\newcommand{\thbar}{{\bar \theta}}
\newcommand{\Dbar}{{\bar D}}

\newcommand{\labar}{{\bar \lambda}}

\def\a{\alpha}
\def\b{\beta}

\def\th{\theta}
\def\g{\gamma}

\def\j{\psi}

\def\F{\Phi}
\def\G{\Gamma}

\def\L{\mathcal{L}}

\def\mon{{{\mathsf{M}}}}

\def\ff#1#2{{\textstyle\frac{#1}{#2}}}

\begin{document}

\allowdisplaybreaks

\renewcommand{\thefootnote}{$\star$}

\renewcommand{\PaperNumber}{068}

\FirstPageHeading

\ShortArticleName{Recent Developments in (0,2) Mirror Symmetry}

\ArticleName{Recent Developments in (0,2) Mirror Symmetry\footnote{This
paper is a contribution to the Special Issue ``Mirror Symmetry and Related Topics''. The full collection is available at \href{http://www.emis.de/journals/SIGMA/mirror_symmetry.html}{http://www.emis.de/journals/SIGMA/mirror\_{}symmetry.html}}}

\Author{Ilarion MELNIKOV~$^\dag$, Savdeep SETHI~$^\ddag$ and Eric SHARPE~$^\S$}

\AuthorNameForHeading{I.~Melnikov, S.~Sethi and E.~Sharpe}

\Address{$^\dag$~Max Planck Institute for Gravitational Physics, Am M\"uhlenberg 1, D-14476 Golm, Germany}
\EmailD{\href{mailto:ilarion.melnikov@aei.mpg.de}{ilarion.melnikov@aei.mpg.de}}

\Address{$^\ddag$~Department of Physics, Enrico Fermi Institute, University of Chicago,\\
\hphantom{$^\ddag$}~5640 S.~Ellis Ave., Chicago, IL  60637, USA}
\EmailD{\href{mailto:sethi@uchicago.edu}{sethi@uchicago.edu}}

\Address{$^\S$~Department of Physics, MC 0435, 910 Drillfield Dr., Virginia Tech,\\
\hphantom{$^\S$}~Blacksburg, VA  24061, USA}
\EmailD{\href{mailto:ersharpe@vt.edu}{ersharpe@vt.edu}}

\ArticleDates{Received June 04, 2012, in f\/inal form October 02, 2012; Published online October 07, 2012}

\Abstract{Mirror symmetry of the type II string has a beautiful generalization to the heterotic string. This generalization, known as (0,2) mirror symmetry, is a f\/ield still largely in its infancy. We describe recent developments including the ideas behind quantum sheaf cohomology, the mirror map for deformations of (2,2) mirrors, the construction of mirror pairs from worldsheet duality, as well as an overview of some of the many open questions. The (0,2) mirrors of Hirzebruch surfaces are presented as a new example.}


\Classification{32L10; 81T20; 14N35}

\renewcommand{\thefootnote}{\arabic{footnote}}
\setcounter{footnote}{0}

\section{Introduction}

In the landscape of string compactif\/ications, the corner comprised of heterotic string compac\-ti\-f\/ications is particularly appealing. Only in this corner is there a possibility of a conventional worldsheet description of f\/lux vacua. In addition, there is a real hope of exploring the interplay between low-energy particle physics and cosmology. These are strong motivations to understand heterotic
worldsheet theories and their associated mathematics more deeply.

For ${\cal N}=1$ space-time supersymmetry, we are interested in worldsheet
theories with (0,2) worldsheet supersymmetry. The special case of models with
(2,2) supersymmetry has been heavily studied in both physics and mathematics.
These are worldsheet theories that can be used to def\/ine type II string
compactif\/ications. Perhaps the most studied and most striking discovery in
(2,2) theories is mirror symmetry: namely, that topologically distinct
target spaces can give rise to isomorphic superconformal f\/ield theories.
Physically, this identif\/ication permits the computation of quantum corrected
observables in one model from protected observables in the mirror model. In
particular, classes of Yukawa couplings can be exactly determined this way.

In mathematics, mirror symmetry has played a crucial role in the development
of curve coun\-ting techniques, quantum cohomology, and modern Gromov--Witten theory. In the the setting of (2,2) models, mirror symmetry is a fairly mature topic. Yet (2,2) theories are a special case of (0,2) models and physically are less
appealing.  Generic ${\cal N}=1$ supersymmetric compactif\/ications of the critical heterotic string,
including the more phenomenologically appealing vacua, are of (0,2) type.
Our aim in this review is to describe recent developments in extending the
ideas of mirror symmetry and quantum cohomology to (0,2) models.

Although general (0,2) superconformal f\/ield theories need not have a geometric
interpretation, there is a familiar geometric set-up that leads to such
theories:  a stable holomorphic bundle~$\cE$ over a smooth Calabi--Yau manifold $X$. The chosen bundle must
satisfy a basic consistency condition to guarantee freedom from anomalies:
\begin{gather}  \label{gs-condition}
{\rm ch}_2({\cal E})   =   {\rm ch}_2(TX).
\end{gather}
In this setting, (0,2) mirror symmetry is the assertion that two topologically
distinct pairs $(X,\cE )$ and $(X^\circ,\cE^\circ)$ of spaces and bundles
can correspond to isomorphic conformal f\/ield theories.  Physically,
having such an
isomorphism can shed light on the structure of the conformal f\/ield theory
and the quantum geometry associated to the classical data $(X,\cE)$.
Mathematically, this isomorphism provides generalizations of
curve counting relations and quantum cohomology.
Much like the (2,2) case, this isomorphism generalizes beyond conformal models
to include massive (0,2) models, including those that describe target spaces with $c_1(T_X) \neq 0$.

The general structure of the correspondence is currently not well understood. However, there is an important special case that has been well-studied by mathematicians and physicists alike. This is the situation where we take the bundle $\cE$ to be the tangent bundle over the Calabi--Yau space. In this case, the resulting conformal f\/ield theory enjoys (2,2) supersymmetry, and (0,2) mirror symmetry reduces to the assertion that a Calabi--Yau~$X$ and its mirror dual~$X^\circ$ lead to isomorphic conformal f\/ield theories. Of course, this is the celebrated (2,2) mirror sym\-metry~\mbox{\cite{Cox:2000vi, Hori:2003ds}}.

There are two ways in which one can seek to generalize the familiar mirror symmetry notions from this starting point. The f\/irst is to consider the (2,2) conformal f\/ield theory on a world-sheet with boundaries. In string theory language, this leads to the study of D-branes on Calabi--Yau manifolds. When we restrict the conformal f\/ield theory data to the topological category (i.e.\ the data associated to certain topological sub-sectors of the full theory), the appropriate mathematical structure is framed by the homological mirror symmetry conjectures~\cite{kont95}.

To describe the second, ``heterotic'' generalization, we observe that the bundle $T_X$ has deformations as a holomorphic bundle over $X$; inf\/initesimally these are counted by $H^1(X, \operatorname{End} T_X )$, and it is easy to f\/ind examples with a large unobstructed moduli space. For instance, in the case of the quintic hypersurface $X \in \mathbb{P}^4$ the tangent bundle has 224 unobstructed deformations. Turning on these deformations reduces the worldsheet supersymmetry to (0,2). As in the D-brane case, one can identify certain (quasi)-topological sub-sectors.
%, where the relevant mathematical structure is that of quantum sheaf cohomology, which will be described in Section~\ref{quantumsheaf}.
In this class of models, (0,2) mirror symmetry then has two primary concerns:
\begin{itemize}\itemsep=0pt
\item how are the deformations of $T_X$ realized in the mirror theory, and what is the map between the two sets of deformations?
\item how does the map relate the quasi-topological observables on the two sides of the mirror?
\end{itemize}
For more general (0,2) theories with bundles $\cE$ not necessarily related to $T_X$ or even of the same rank, we need to ask more basic questions like:
\begin{itemize}\itemsep=0pt
\item how do we characterize mirror pairs?
\item how do we compute the non-perturbative ef\/fects which give (0,2) generalizations
of curve counting and quantum cohomology?
\end{itemize}
We should stress that given an isomorphism of the heterotic conformal f\/ield theories for $(X,T_X)$ and $(X^\circ,T_X^\circ)$, as well as the existence of unobstructed deformations for the $(X,T_X)$ conformal f\/ield theory, we know on general grounds that corresponding deformations must exist on the mirror side, and that the deformed theories must remain isomorphic as (0,2) theories.  This is the crucial point that makes the ``(2,2) locus'' (i.e.\ the choice $\cE = T_X$)  a natural starting point for explorations of (0,2) mirror symmetry:  we are assured of success, and our primary job is to f\/ind the appropriate map.

Just as ordinary mirror symmetry exchanges the topological A and B models,
(0,2) mirror symmetry exchanges what are termed the A/2 and B/2 models.
We therefore begin in Section~\ref{quantumsheaf} with a discussion of
the A/2 model.  If $X$ and $X^{\circ}$ are (2,2) mirror Calabi--Yau spaces then their Hodge numbers are
exchanged,
\[
h^{i,j}(X)   =   h^{n-i,j}(X^{\circ}),
\]
where $n$ is the dimension of both $X$  and $X^{\circ}$. Instead of exchanging Hodge numbers, sheaf cohomology groups are exchanged for (0,2) mirror pairs:
\begin{gather}  \label{sc-exchange}
h^j\left(X, \wedge^i {\cal E}^* \right)   =
h^j\left(X^{\circ}, \wedge^i {\cal E}^{\circ} \right).
\end{gather}
Non-perturbative ef\/fects are therefore encoded in ``quantum sheaf
cohomology'', which
gene\-ra\-lizes the  quantum cohomology ring of (2,2) mirror symmetry.  This is physically a ground ring, isomorphic to a deformation of a classical cohomology ring, which is corrected by non-perturbative ef\/fects. In Section~\ref{quantumsheaf}, we review the current status of
quantum sheaf cohomology.

One of the early successes of
the mirror symmetry program was f\/inding a precise
map between complex and K\"ahler moduli for certain mirror pairs,
known as the monomial-divisor mirror map~\cite{Aspinwall:1993rj}.
If one considers deformations of $T_X$ for Calabi--Yau
spaces def\/ined by ref\/lexively plain polytopes then a (0,2) generalization of the monomial-divisor mirror map
exists~\cite{Melnikov:2010sa}. The map describes the exchange of complex, K\"ahler and
bundle moduli. Even in this class of models, f\/inding the map is relatively challenging, and the story is far from complete.  However, the last few years has seen some important progress, leading to interesting structures and opening up new routes for further investigation. We describe this progress in Section~\ref{deformtangent}.

In Section~\ref{worldsheetduality}, we describe the construction of mirror pairs using worldsheet duality~\cite{Adams:2003d}. This generalizes the physical approach taken in~\cite{Hori:2000kt}\ to construct (2,2) mirror pairs for non-compact toric spaces. In the (0,2) setting, worldsheet duality leads to mirror descriptions of sigma models with toric target spaces, including non-compact Calabi--Yau spaces and models without a (2,2) locus. As an example, we present the (0,2) mirrors for Hirzebruch surfaces.

In terms of historical development, explorations of (0,2) mirror symmetry go back more than a decade. Early work provided numerical evidence for the existence
of a (0,2) duality by computing the dimensions of sheaf cohomology groups in
a large class of examples~\cite{bsw}. A visual check showed that most cases came in pairs satisfying the symmetry~(\ref{sc-exchange}). Later work generalized the original Greene--Plesser construction~\cite{gp1}  to a class of~(0,2) models, with mirrors generated via orbifolding by a f\/inite symmetry group~\cite{Blumenhagen:1997pp, blum-s}.

The more recent work was initiated by the construction of (0,2) mirrors via worldsheet duali\-ty~\cite{Adams:2003d}, which led to precise def\/initions of heterotic chiral rings~\cite{Adams:2003d, Adams:2005tc}, and to the notion of quantum sheaf cohomology~\cite{Donagi:2011uz, Donagi:2011, Katz:2004nn, s3, s2}. Most of our presentation of mirror maps for deformations of $T_X$ is based on~\cite{Kreuzer:2010ph,McOrist:2008ji,Melnikov:2010sa}. Perhaps the most important point to stress is how many of the central questions remain wide open. The f\/ield of (0,2) mirror symmetry and (0,2) string compactif\/ications is still truly in its infancy. In Section~\ref{summary}, we end by overviewing some of the current and future directions of investigation.

\section{Quantum sheaf cohomology}
\label{quantumsheaf}

We begin by describing computations of non-perturbative
corrections in (0,2) theories, which are encapsulated in the notion of
``quantum sheaf cohomology''.  This
is a generalization of ordinary quantum cohomology.
Instead of giving a quantum deformation of ordinary cohomology rings,
quantum sheaf cohomology gives a deformation of sheaf cohomology rings.
Specif\/ically, let $X$ be a complex K\"ahler manifold and ${\cal E}
\rightarrow X$ a holomorphic vector bundle satisfying the following two
conditions:
\begin{gather}\label{qcconditions}
\wedge^{\rm top} {\cal E}^*   \cong   K_X,   \qquad
{\rm ch}_2({\cal E})   =   {\rm ch}_2(TX).
\end{gather}
A pair $(X, {\cal E})$ satisfying the conditions~\eqref{qcconditions}
is sometimes known as ``omalous'', which is a shorte\-ning of ``non-anomalous''.
These conditions are slightly stronger than the basic conditions needed for
a consistent (0,2) theory, and arise from demanding that the A/2 theory, to be reviewed shortly,
be well-def\/ined.

Quantum sheaf cohomology is then a deformation of the classical ring
generated by
\[
H^*(X, \wedge^* {\cal E}^*).
\]
In the special case that ${\cal E} = TX$, quantum sheaf cohomology reduces
to ordinary quantum cohomology.

Historically, the existence of quantum sheaf cohomology was f\/irst proposed
in~\cite{Adams:2003d}.  The paper~\cite{Katz:2004nn} worked out suf\/f\/icient
details to mathematically compute in examples.  Questions concerning
existence of OPE ring structures in theories with only (0,2) supersymmetry
were discussed in~\cite{Adams:2005tc}.  The subject has since been further
developed in a number of works inclu\-ding~\cite{Borisov:2011xu,Donagi:2011uz,Donagi:2011,Guffin:2011mx,Guffin:2007mp,Kreuzer:2010ph,McOrist:2010ae,McOrist:2007kp,mm3,
McOrist:2008ji,m1,ms,s3,s2,tan1,tan2,
Tan:2008mi,
Yagi:2010tp}.

In principle, quantum sheaf cohomology should exist for any omalous pair
$(X, {\cal E})$; that said, at the moment, computational techniques only
exist in more limited cases.  Specif\/ically,
the special case that $X$ is a toric variety and ${\cal E}$ is
a deformation of the tangent bundle is well-understood. For hypersurfaces
in toric varieties, there is a ``quantum restriction'' proposal
\cite{McOrist:2008ji} that generalizes an old technique of Kontsevich, though
more work on hypersurfaces is certainly desirable.

{\sloppy Ordinary quantum cohomology arises physically as a description of
correlation functions in the A model topological f\/ield theory,
\begin{gather*}
S_A= \frac{1}{\alpha'} \int_{\Sigma} d^2z \left(
         \left( g_{\mu \nu} +  i B_{\mu \nu} \right)
\partial \phi^{\mu} \overline{\partial} \phi^{\nu}\vphantom{\frac{i}{2}}\right.\\
\left.\hphantom{S_A=}{}  +
\frac{i}{2} g_{\mu \nu} \psi_+^{\mu} D_{\overline{z}} \psi_+^{\nu}
 +  \frac{i}{2} g_{\mu \nu} \psi_-^{\mu} D_z \psi_-^{\nu}
 +  R_{i \overline{\jmath} k \overline{l}}
\psi_+^i \psi_+^{\overline{\jmath}} \psi_-^k \psi_-^{\overline{l}}
\right),
\end{gather*}
where $\phi: \Sigma \rightarrow X$ is a map from the worldsheet~$\Sigma$
into the space $X$ in which the string propagates, and the $\psi_{\pm}^\mu$ are
fermionic superpartners of the coordinates~$\phi^\mu$ on~$X$.
There is a nilpotent scalar operator~$Q$, known as the
BRST operator, whose action on the f\/ields above is schematically as follows:
\begin{gather*}
\delta \phi^i  \propto  \chi^i,  \qquad
\delta \phi^{\overline{\imath}}   \propto   \chi^{\overline{\imath}},
\qquad
\delta \chi^i   =   0   =   \delta \chi^{\overline{\imath}}, \qquad
\delta \psi^{\overline{\imath}}_{\overline{z}}   \neq   0, \qquad
\delta \psi^{i}_z   \neq   0.
\end{gather*}
The states of the theory are BRST-closed dimension zero operators
(modulo BRST-exact ope\-ra\-tors),
which constrains them to be of the form
\begin{gather}\label{amodelstates}
b_{i_1 \dots i_p \overline{\imath}_1 \dots \overline{\imath}_q}(\phi)
\chi^{i_1} \cdots \chi^{i_p} \chi^{\overline{\imath}_1} \cdots
\chi^{\overline{\imath}_q}.
\end{gather}
Witten
observed that the states of~\eqref{amodelstates} are in one-to-one correspondence with
dif\/ferential forms~\cite{edtft},
\begin{gather*}
b_{i_1 \dots i_p \overline{\imath}_1 \dots \overline{\imath}_q}(\phi)
dz^{i_1} \wedge \cdots \wedge dz^{i_p} \wedge
d \overline{z}^{\overline{\imath}_1} \wedge \cdots \wedge
d \overline{z}^{\overline{\imath}_q},
\end{gather*}
and the BRST operator $Q$ with the exterior derivative $d$,
hence the states are in one-to-one correspondence with elements of
$H^{p,q}(X)$.

}

The A/2 model is def\/ined by the action:
\begin{gather*}
 S_{A/2}=\frac{1}{\alpha'} \int_{\Sigma} d^2z \left(
\left(g_{\mu \nu} +  i B_{\mu \nu} \right)
\partial \phi^{\mu} \overline{\partial} \phi^{\nu} \vphantom{\frac{i}{2}} \right.\\
\left. \hphantom{S_{A/2}=}{} +   \frac{i}{2} g_{\mu \nu} \psi_+^{\mu} D_{\overline{z}} \psi_+^{\nu}
  +   \frac{i}{2}h_{\alpha \beta} \lambda_-^{\alpha} D_{z} \lambda_-^{\beta}
 +
F_{i \overline{\jmath} a \overline{b}} \psi_+^i \psi_+^{\overline{\jmath}}
\lambda_-^a \lambda_-^{\overline{b}}
\right).
\end{gather*}
In the special case of ${\mathcal E} = TX$, the A/2 model becomes
the A model.
Anomaly cancellation in the A/2 model requires
\begin{gather*}
\wedge^{\rm top} {\mathcal E}^*   \cong   K_X, \qquad
{\rm ch}_2({\mathcal E})   =   {\rm ch}_2(TX).
\end{gather*}
The f\/irst statement is a condition specif\/ic to the A/2 model,
an analogue of the condition that the closed string B model can only propagate
on spaces $X$ such that $K_X^{\otimes 2} \cong {\cal O}_X$
\cite{s3,edtft}.  The second statement is commonly known as the ``Green--Schwarz
anomaly cancellation condition'', and is generic to all heterotic theories.

There is a BRST operator $Q$ in the A/2 model, which acts as follows:
\begin{gather*}
\delta \phi^i   =   0, \qquad
\delta \phi^{\overline{\imath}}  \propto  \psi_+^{\overline{\imath}}, \qquad
 \delta \psi_+^{\overline{\imath}}   =   0   =
\delta \lambda_-^a, \qquad
\delta \psi_+^i   \neq   0, \qquad
\delta \lambda_-^{\overline{a}}   \neq   0.
\end{gather*}
The states of the A/2 model generalizing the A model states
are of the form
\begin{gather}\label{a2obs}
b_{\overline{\imath}_1 \dots \overline{\imath}_q a_1 \dots a_p}
\psi_+^{\overline{\imath}_1} \cdots \psi_+^{\overline{\imath}_q}
\lambda_-^{a_1} \cdots \lambda_-^{a_p}.
\end{gather}
Proceeding in an analogous fashion, we identify the BRST operator $Q$
with $\overline{\partial}$, and the states of~\eqref{a2obs} with elements of
sheaf cohomology $H^q(X, \wedge^p \mathcal{E}^{*})$.

In addition to the A/2 model which provides the (0,2) version of the A model,
there also exists a B/2 model which provides a (0,2) version of the B model
topological f\/ield theory.  As one might guess, (0,2) mirror symmetry exchanges
A/2 and B/2 model correlation functions, just as ordinary mirror symmetry
exchanges A and B model correlation functions.  There are also some
surprising new symmetries; for example, the B/2 model on~$X$ with bundle
${\cal E}$ is equivalent, at least classically, to the A/2 model on~$X$
with bundle ${\cal E}^*$.  We do not have space here to discuss the
B/2 model separately (and indeed, given the symmetry just mentioned, little
discussion is really needed); see instead~\cite{s3} for further
information.

\subsection{Formal computations}

Quantum sheaf cohomology is determined by correlation function computations
in the A/2 model.  In this section we will brief\/ly outline how such
computations are def\/ined, at least at a~formal level.

First consider the case with no $(\psi_+^{\overline{\imath}}$, $\lambda_-^a)$
zero modes (no ``excess intersection''). In this case, a correlation function
will have the form,
\begin{gather}\label{correlator}
\langle {\mathcal O}_1 \cdots {\mathcal O}_n \rangle   =
\sum_{\beta} \int_{{\mathcal M}_{\beta}} \omega_1 \wedge \cdots\wedge
\omega_n,
\end{gather}
where ${\cal M}_{\beta}$ is a moduli space of curves of degree $\beta$,
and $\omega_i$ is an element of
$H^q({\mathcal M}_{\beta}, \wedge^p {\mathcal F}^*)$
induced by the element of $H^q(X, \wedge^p {\mathcal E}^*)$ corresponding
to~${\mathcal O}_i$.
The sheaf ${\mathcal F}$ is induced from~${\mathcal E}$.  For example,
if the moduli space ${\mathcal M}$ admits a universal instanton $\alpha$,
then ${\mathcal F} = R^0 \pi_* \alpha^* {\mathcal E}$.

The integrand above is an element of
\begin{gather*}
H^{\sum q_i}\big( {\mathcal M}_{\beta},
\wedge^{\sum p_i} {\mathcal F}^* \big),
\end{gather*}
and so will vanish unless
\begin{gather*}
\sum q_i   =   \dim {\mathcal M}_{\beta}, \qquad
\sum p_i   =   \operatorname{rank} {\mathcal F}.
\end{gather*}
Furthermore,
Grothendieck--Riemann--Roch tells us that the conditions for $(X,
{\mathcal E})$ to be omalous, namely
\begin{gather*}
\wedge^{\rm top} {\mathcal E}^* \cong K_X, \qquad
{\rm ch}_2({\mathcal E}) = {\rm ch}_2(TX),
\end{gather*}
imply that, at least formally, $\wedge^{\rm top} {\mathcal F}^* \cong
K_{{\mathcal M}_{\beta}}$, which guarantees that the integrand of~\eqref{correlator} determines
a number.

Next, let us brief\/ly consider the case of excess intersection where there are $(\psi_+^{\overline{\imath}},\lambda_-^a)$ zero modes.
In the ordinary A model, this would involve adding an Euler class factor,
corresponding physically to using four-fermi terms to soak up extra
zero modes.  In the A/2 model, four-fermi terms are interpreted as
generating elements of
\begin{gather*}
H^1\left( \mathcal{M}_{\beta}, \mathcal{F}^* \otimes \mathcal{F}_1 \otimes
\left( \mbox{Obs} \right)^* \right).
\end{gather*}
Taking into account those four-fermi terms, the integrand is then an
element of
\begin{gather*}
H^{\rm top}\left( {\mathcal M}_{\beta}, \wedge^{\rm top} {\mathcal F}^* \otimes
\wedge^{\rm top} {\mathcal F}_1 \otimes \wedge^{\rm top} {\rm Obs}^* \right).
\end{gather*}
As before, Grothendieck--Riemann--Roch and the anomaly cancellation conditions
imply that the integrand determines a number.

\subsection{Linear sigma model compactif\/ications}

In order to do actual computations, we need to pick a compactif\/ication
of the moduli space of curves, and describe how to extend the induced sheaves
${\cal F}$, ${\cal F}_1$ over that compactif\/ication.
We will work with toric varieties and linear sigma model compactif\/ications
of moduli spaces.  These compactif\/ications are well-known, so we shall
be brief.

Schematically, a linear sigma model compactif\/ication of a moduli space of
curves in a toric variety is built by expressing the toric variety
as a $\mathbb{C}^{\times}$ quotient, expanding the homogeneous coordinates
in zero modes, and then taking those zero modes to be homogeneous
coordinates on the moduli space (with the same $\mathbb{C}^{\times}$ quotient
and weightings as the original homogeneous coordinates, and exceptional
set determined by that of the original space).
For example, consider~${\mathbb P}^{N-1}$, which is described as a
${\mathbb C}^{\times}$ quotient of $N$ homogeneous coordinates, each of
weight~1.  For a moduli space of maps ${\mathbb P}^1 \rightarrow
\mathbb{P}^{N-1}$ of degree $d$, we expand each homogeneous coordinate
in a basis of sections of $\phi^* {\cal O}(1) = {\cal O}(d)$, and interpret
coef\/f\/icients as homogeneous coordinates on the moduli space.
Since the space of sections of ${\cal O}(d)$ has dimension $d+1$,
that means the moduli space is a ${\mathbb C}^{\times}$ quotient of
${\bf C}^{N(d+1)}$, which naturally leads to ${\mathbb P}^{N(d+1)-1}$.

The construction of induced sheaves is described in
\cite{Donagi:2011uz,Donagi:2011,Katz:2004nn}.
Schematically, it works as follows: in the original physical theory,
the bundle ${\cal E}$ is built from kernels and cokernels of maps between
direct sums of line bundles, i.e.\ sums of powers of the universal
sub-bundle $S$ and analogues thereof.  Brief\/ly, we lift each such line bundle
on the original toric variety to a line bundle on
${\mathbb P}^1 \times {\cal M}$, in such a way that sums and powers of
universal sub-bundles are preserved.  After lifting, these are then
pushed forward to~${\cal M}$.

For example, consider the completely reducible bundle,
\begin{gather*}
{\mathcal E}   =   \oplus_a {\mathcal O}(n_a),
\end{gather*}
on $\mathbb{P}^{N-1}$.
Corresponding to the universal sub-bundle,
\begin{gather*}
S := {\mathcal O}(-1)   \longrightarrow   \mathbb{P}^{N-1},
\end{gather*}
is
\begin{gather*}
{\mathcal S}   =   \pi_1^* {\mathcal O}_{\mathbb{P}^1}(-d) \otimes
\pi_2^* {\mathcal O}_{\mathcal M}(-1)   \longrightarrow
\mathbb{P}^1 \times {\mathcal M}.
\end{gather*}
The lift of ${\mathcal E}$ is
\begin{gather*}
\oplus_a {\mathcal S}^{\otimes -n_a}   \longrightarrow
\mathbb{P}^1 \times {\mathcal M},
\end{gather*}
which pushes forward to
\begin{gather*}
{\mathcal F}   =   \oplus_a H^0\left( \mathbb{P}^1, {\mathcal O}(n_a d)
\right) \otimes_{\mathbb{C}} {\mathcal O}(n_a), \qquad
{\mathcal F}_1   =   \oplus_a H^1\left( \mathbb{P}^1, {\mathcal O}(n_a d)
\right) \otimes_{\mathbb{C}} {\mathcal O}(n_a).
\end{gather*}
This generalizes to other toric varieties as well as to Grassmannians.
Physically, this is equivalent to expanding worldsheet fermions in
a basis of zero modes, and identifying each basis element with a line
bundle of the same~$\mathbb{C}^{\times}$ weights as the original
line bundle.

The example above illustrates what happens for gauge bundles that are
sums of line bundles.  Next, let us consider a cokernel of a map between
sums of line bundles:
\begin{gather*}
0   \longrightarrow   {\mathcal O}^{\oplus k}   \longrightarrow
\oplus_i {\mathcal O}(\vec{q}_i)   \longrightarrow   {\mathcal E}
  \longrightarrow   0,
\end{gather*}
over some toric variety~$X$.
Lifting to ${\mathbb P}^1 \times {\cal M}$ and pushing forward gives the
long exact sequence
\begin{gather*}
0   \longrightarrow   \oplus_k H^0({\mathcal O}) \otimes {\mathcal O}
  \longrightarrow   \oplus_i H^0({\mathcal O}(\vec{q}_i \cdot \vec{d}))
\otimes {\mathcal O}(\vec{q}_i)
  \longrightarrow   {\mathcal F} \\
 \hphantom{0}{} \longrightarrow   \oplus_k H^1({\mathcal O}) \otimes {\mathcal O}
  \longrightarrow   \oplus_i H^1({\mathcal O}(\vec{q}_i \cdot \vec{d}))
\otimes {\mathcal O}(\vec{q}_i)
  \longrightarrow   {\mathcal F}_1   \longrightarrow   0,
\end{gather*}
which simplif\/ies to the statements
\begin{gather*}
0   \longrightarrow   {\mathcal O}^{\oplus k}   \longrightarrow
\oplus_i H^0({\mathcal O}(\vec{q}_i \cdot \vec{d}))
\otimes {\mathcal O}(\vec{q}_i)
  \longrightarrow   {\mathcal F}   \longrightarrow   0,
\\
{\mathcal F}_1   \cong   \oplus_i H^1({\mathcal O}(\vec{q}_i \cdot \vec{d}))
\otimes {\mathcal O}(\vec{q}_i).
\end{gather*}
It can be shown that if ${\mathcal E}$ is locally-free,
then ${\mathcal F}$ will also be locally-free.

As a consistency check, let us examine the special case that
${\mathcal E} = TX$.  The tangent bundle of a (compact, smooth) toric
variety can be expressed as a cokernel
\begin{gather*}
0   \longrightarrow   {\mathcal O}^{\oplus k}   \longrightarrow
\oplus_i {\mathcal O}(\vec{q}_i)   \longrightarrow   TX
  \longrightarrow   0.
\end{gather*}
Applying the previous ansatz, we have
\begin{gather*}
0   \longrightarrow   {\mathcal O}^{\oplus k}   \longrightarrow
\oplus_i H^0({\mathcal O}(\vec{q}_i \cdot \vec{d}))
\otimes {\mathcal O}(\vec{q}_i)
  \longrightarrow   {\mathcal F}   \longrightarrow   0,
\\
{\mathcal F}_1   \cong   \oplus_i H^1({\mathcal O}(\vec{q}_i \cdot \vec{d}))
\otimes {\mathcal O}(\vec{q}_i).
\end{gather*}
In this case, for ${\mathcal E}=TX$, we expect ${\mathcal F} = T {\mathcal M}$
and ${\mathcal F}_1$ to be the obstruction sheaf in the sense of~\cite{am,Morrison:1994fr}.  The sequences above have precisely the right form,
and it can be shown that the induced maps are also correct.

\subsection{Results}

Consider a deformation of the tangent bundle of a toric variety $X$ def\/ined
as a cokernel
\begin{gather*}
   0   \longrightarrow   {\cal O}_X \otimes_\mathbb{C} W^*
\mathop{\longrightarrow}^{E}
   {\bigoplus}_\rho
   {\cal O}_X(D_\rho)   \longrightarrow   {\cal E}
\longrightarrow   0,
   %\label{eq:Epresentation}
\end{gather*}
where $D_{\rho}$  parametrizes the toric divisors, and $W = H^2(X, {\mathbb C})$.
The components of $E$ are a~collection of $W$-valued sections $E_{\rho}$
of ${\cal O}_X(D_{\rho})$.  We write,
\begin{gather*}
E_{\rho}   =   \sum_{\rho'} a_{\rho \rho'} x_{\rho'}   +   \cdots,
\end{gather*}
where $a_{\rho \rho'} \in W$, $x_{\rho'}$ is the homogeneous coordinate in the
Cox ring associated to $\rho'$ (i.e.\  $x_{\rho} \in
H^0(X, {\cal O}_X(D_{\rho}))$).  Nonlinear terms in $x$'s have been omitted.

For each linear equivalence class $c$ of toric divisors,
def\/ine a matrix $A_c$ to be the $|c| \times |c|$ matrix whose entries are
$a_{\rho \rho'}$, where $\rho$, $\rho'$ are within the same linear
equivalence class $c$.  Def\/ine:
\[ Q_c = \det A_c.
\]
Recall that a collection of edges $K$ of the fan is called a primitive collection
if $K$ does not span any cone in the fan, but every proper sub-collection
of $K$ does.  Equivalently, the intersection of all of the divisors in $K$
is empty, but the intersection of any sub-collection is non-empty.

It was shown in \cite{Donagi:2011uz} that any primitive collection $K$
is a union
of linear equivalence classes.  With that in mind, def\/ine $Q_K$ to be
the product of all~$Q_c$ for~$c$ a linear equivalence class contained in $K$.

Def\/ine:
\[
SR(X,{\cal E})   =   \{ Q_K \, | \, \mbox{$K$~a primitive collection} \}.
\]
It was shown in~\cite{Donagi:2011uz}\ that the classical product
structure on $\oplus H^*(X, \wedge^* {\cal E}^*)$ is encoded in the statement,
\begin{gather*}
\oplus H^*(X, \wedge^* {\cal E}^*)   =   {\rm Sym}^* W / SR(X, {\cal E}).
\end{gather*}
Because all linear sigma model moduli spaces for toric varieties are also toric,
the classical sheaf cohomology ring in each separate
worldsheet instanton sector has the same form:
\begin{gather*}
\oplus H^*\left({\cal M}_{\beta}, \wedge^* {\cal F}_{\beta}^* \right)
  =
{\rm Sym}^* W / \widehat{SR}\left( {\cal M}_{\beta}, {\cal F}_{\beta} \right),
\end{gather*}
where
\begin{gather*}
\widehat{SR}\left( {\cal M}_{\beta}, {\cal F}_{\beta} \right)   =
\{ Q_{K_{\beta}} \, | \,\mbox{$K$ a primitive collection} \}
\end{gather*}
and
\begin{gather*}
Q_{K_{\beta}}   =   \prod_{c \in [K]} Q_c^{h^0(D_c \cdot \beta)},
\end{gather*}
where $[K]$ denotes the set of linear equivalence classes of the
$D_{\rho}$ with $\rho \in K$, for $K$ a primitive collection
($D_c \cdot \beta$ means $D_{\rho} \cdot \beta$ for any $\rho$ in the
linear equivalence class~$c$).

To generate the quantum sheaf cohomology relations, one must f\/ind relations
between correlation functions in dif\/ferent instanton sectors.
For the sake of brevity, we omit the details here,
and instead refer the interested reader
to~\cite{Donagi:2011uz,Donagi:2011}.

To def\/ine the f\/inal result, we must def\/ine a set $K^-$, and a class
$\beta_K \in H_2(X,{\mathbb Z})$.
For any primitive collection $K$ consider the element,
\begin{gather*}
v   =   \sum_{\rho \in K} v_{\rho},
\end{gather*}
of the toric lattice.  Then $v$ lies in the relative interior of a unique
cone $\sigma$.  Let $K^-$ denote the set of edges of $\sigma$.
Then one can write
\begin{gather*}
v   =   \sum_{\rho \in K^-} c_{\rho} v_{\rho},
\end{gather*}
where each $c_{\rho} > 0$, hence
\begin{gather*}
\sum_{\rho \in K} v_{\rho}   =   \sum_{\rho \in K^-} c_{\rho} v_{\rho},
\end{gather*}
or equivalently
\begin{gather}  \label{beta-kernel}
\sum a_{\rho} v_{\rho}   =   0.
\end{gather}
Dualizing the sequence
\begin{gather*}
0   \longrightarrow   M   \longrightarrow   {\mathbb Z}^{\Sigma(1)}
  \longrightarrow   {\rm Pic}(X)   \longrightarrow   0,
\end{gather*}
we see that equation~(\ref{beta-kernel}) can be induced by intersection
with elements of $H_2(X,{\mathbb Z})$.  Hence, for each primitive collection
$K$, there is a class $\beta_K \in H_2(X,{\mathbb Z})$ such that
$D_{\rho} \cdot \beta_K = a_{\rho}$; see also~\cite{Batyrev:1993,Donagi:2011uz} for more information on this notation.

Finally,~\cite{Donagi:2011uz,Donagi:2011}\ show that the quantum
sheaf cohomology relations are given by
\begin{gather*}
\prod_{c \in [K]} Q_c   =
q^{\beta_K}
\prod_{c \in [K^-]} Q_c^{- D_c \cdot \beta_K},
\end{gather*}
for quantum parameters $q^{\beta_K}$.

In this section we have outlined a mathematical description of quantum sheaf
cohomology.  The same results (at least for purely linear maps $E$) can
also be obtained physically from one-loop ef\/fective action arguments on
Coulomb branches of gauged linear sigma models, in analogy with
\cite{Morrison:1994fr}; see, for example,~\cite{McOrist:2007kp,McOrist:2008ji}.

\subsection[Example: $\mathbb{P}^1 \times \mathbb{P}^1$]{Example: $\boldsymbol{\mathbb{P}^1 \times \mathbb{P}^1}$}

Let $X = {\mathbb P}^1 \times {\mathbb P}^1$, and
consider the vector bundle ${\cal E}$ given as the cokernel
\begin{gather*}
0   \longrightarrow   {\cal O} \oplus {\cal O}   \stackrel{E}{
\longrightarrow}
{\cal O}(1,0)^2 \oplus {\cal O}(0,1)^2   \longrightarrow   {\cal E}
  \longrightarrow   0,
\end{gather*}
where
\begin{gather*}
E   =  \left[ \begin{array}{@{}cc@{}}
Ax & Bx \\
C \tilde{x} & D \tilde{x}   \end{array} \right],
\end{gather*}
where $A$, $B$, $C$, $D$ are $2 \times 2$ matrices and
\begin{gather*}
x   =   \left[ \begin{array}{@{}c@{}} x_1 \\ x_2 \end{array} \right],       \quad
\tilde{x}   =   \left[ \begin{array}{@{}c@{}} \tilde{x}_1 \\ \tilde{x}_2
\end{array} \right]
\end{gather*}
are arrays of homogeneous coordinates on the two ${\mathbb P}^1$ factors.
The bundle ${\cal E}$ is a deformation of the tangent bundle, which itself corresponds
to the special case $A = D = I_{2 \times 2}$, $B = C = 0$). In general $\cE$
is not isomorphic to the tangent bundle.
It can be shown that $H^1(X, {\cal E}^*)$ is two-dimensional.

In the language of the previous section, there are two primitive
collections. The two $Q_K$'s corresponding to each of those primitive
collections are
\begin{gather*}
\det\big( A \psi + B \tilde{\psi} \big),      \qquad
\det\big( C \psi + D \tilde{\psi} \big).
\end{gather*}
For both of those primitive collections, $K^-$ is empty.

It can be shown \cite{Donagi:2011uz,Donagi:2011,McOrist:2007kp,McOrist:2008ji}
that the
quantum sheaf cohomology ring
for this case is given by
${\mathbb C}[\psi, \tilde{\psi}]$ modulo the relations,
\begin{gather*}
\det\big( A \psi + B \tilde{\psi} \big)   =   q,       \qquad
\det\big( C \psi + D \tilde{\psi} \big)   =   \tilde{q},
\end{gather*}
where $\psi$, $\tilde{\psi}$ form a basis for $H^1(X, {\cal E}^*)$.
As a consistency check, note that in the special case that ${\cal E} = TX$,
the relations above reduce to
\begin{gather*}
\psi^2   =   q,\qquad
\tilde{\psi}^2   =   \tilde{q}
\end{gather*}
duplicating the ordinary quantum cohomology ring of ${\mathbb P}^1 \times
{\mathbb P}^1$.

\subsection{General Hirzebruch surfaces}
\label{mathhirze}

Let us now outline a similar computation for more general Hirzebruch surfaces ${\mathbb F}_n$.
Describe such a surface as a quotient of
the homogeneous coordinates $u$, $v$, $s$, $t$ (corresponding to the
four toric divisors) by two ${\bf C}^{\times}$
actions with the following weights:
\[
\begin{array}{cccc}
u & v & s & t \\ \hline
1 & 1 & 0 & n \\
0 & 0 & 1 & 1
\end{array}
\]
Without loss of generality, we assume $n \geq 0$.
Describe a deformation of the tangent bundle as the cokernel
\begin{gather*}
0   \longrightarrow   {\cal O} \oplus {\cal O}
\stackrel{E}{\longrightarrow}
{\cal O}(1,0)^{\oplus 2} \oplus {\cal O}(0,1) \oplus {\cal O}(n,1)
  \longrightarrow   {\cal E}   \longrightarrow   0,
\end{gather*}
where
\begin{gather*}
E   =   \left[ \begin{array}{@{}cc@{}}
A x & B x \\
\gamma_1 s & \gamma_2 s \\
\alpha_1 t + s f_1(u,v) & \alpha_2 t + s f_2(u,v)
\end{array} \right],
\end{gather*}
with
\begin{gather*}
x   \equiv   \left[ \begin{array}{@{}c@{}} u \\ v \end{array} \right],
\end{gather*}
$A$, $B$ constant $2 \times 2$ matrices, $\gamma_1$, $\gamma_2$,
$\alpha_1$, $\alpha_2$ constants, and $f_{1,2}(u,v)$ homogeneous polynomials
of degree~$n$.
In particular when $n > 0$, the polynomials $f_i$ def\/ine ``nonlinear''
deformations, in the sense that they def\/ine nonlinear entries in~$E$.

Let us def\/ine
\begin{gather*}
Q_{K1}  =  \det \big( \psi A   +   \tilde{\psi} B \big), \qquad
Q_s  =   \psi \gamma_1   +   \tilde{\psi} \gamma_2, \qquad
Q_t  =  \psi \alpha_1   +   \tilde{\psi} \alpha_2,
\end{gather*}
following the nomenclature used in~\cite{Donagi:2011uz, Donagi:2011}.
The Hirzebruch surface has two primitive collections, with corresponding
$Q$'s: $Q_{K1}$ and $Q_s Q_t$.
Following the general procedure described earlier,
the quantum sheaf cohomology ring is given by ${\mathbb C}[\psi,
\tilde{\psi}]$ modulo the relations,
\[
Q_{K1}   =   q_1 Q_s^n,       \qquad
Q_s Q_t   =   q_2.
\]

Although the quantum sheaf cohomology relations above depend upon
the ``linear'' para\-me\-ters $(A, B,  \gamma_{1,2},   \alpha_{1,2})$,
they do not depend on the nonlinear contributions from the polyno\-mials~$f_{1,2}(u,v)$.  In fact, this is a general feature of these toric
quantum sheaf cohomology relations; namely that nonlinear deformations
drop out of the quantum sheaf cohomology.

For completeness, let us also consider the special case where
${\cal E} = TX$ which is the (2,2) locus.
This special case is described by
\begin{gather*}
A   =   I,       \qquad B   =   0,       \qquad
\gamma_1   =   0, \qquad       \gamma_2   =   1,       \qquad
\alpha_1   =   n, \qquad       \alpha_2   =   1,       \qquad
f_1   =   f_2   =   0,
\end{gather*}
so that
\begin{gather*}
Q_{K1}   =   \psi^2, \qquad
Q_s   =   \tilde{\psi},   \qquad    Q_t   =   n \psi   +   \tilde{\psi}.
\end{gather*}
In this special case, if we identify
$D_u = \psi$, $D_s = \tilde{\psi}$, then
the quantum sheaf cohomology ring reduces to
\begin{gather*}
D_u^2   =   q_1 D_s^n,       \qquad
D_s( n D_u   +   D_s)   =   q_2,
\end{gather*}
which, for example, reduces to the classical cohomology ring relations
when $q_1, q_2 \rightarrow 0$.

\section{Mirror maps}
\label{deformtangent}

In the preceding section, we described the A/2 model and the (0,2) analogue
of curve counting encoded in quantum sheaf cohomology.  With those preliminaries in place, we turn to the
mirror map for deformations of the tangent bundle.

Since (2,2) mirror symmetry is a basic tool in our current
understanding of more general (0,2)  phenomena, we begin with a
brief review of the (2,2) case.  As we will see, many of the subtleties of
the (0,2) story are already familiar from the better-understood (2,2) setting;
however, there are also important complications that are intrinsically (0,2) in nature.

\subsection{(2,2) mirror symmetry \`a la Batyrev}

We begin by reviewing what is
perhaps the best-known general construction of mirror pairs~$X$ and~$X^\circ$~\cite{Batyrev:1994hm,Cox:2000vi}.  Consider a~$d$-dimensional lattice polytope $\Delta$ containing the origin in a~$d$-dimensional lattice $\bM \subset \bM_{\R} \sim \R^d$.  Let $\bN \subset \bN_{\R}$ be the dual lattice, and denote the natural pairing $\bM_{\R} \times \bN_{\R}$ by $\langle\cdot,\cdot\rangle$.
The dual polytope $\Delta^\circ \subset \bN_{\R}$ is def\/ined by
\[
\Delta^\circ = \{ y \in \bN_{\R} \,|\,  \langle x,y \rangle \ge -1 \  \forall\,
x \in \Delta\}.
\]
$\Delta$ is ref\/lexive if\/f $\Delta^\circ$ is also a lattice polytope; note that $\Delta^\circ$ is ref\/lexive if\/f $\Delta$ is ref\/lexive.
A~familiar $d=2$ example is given in~(\ref{eq:P2})
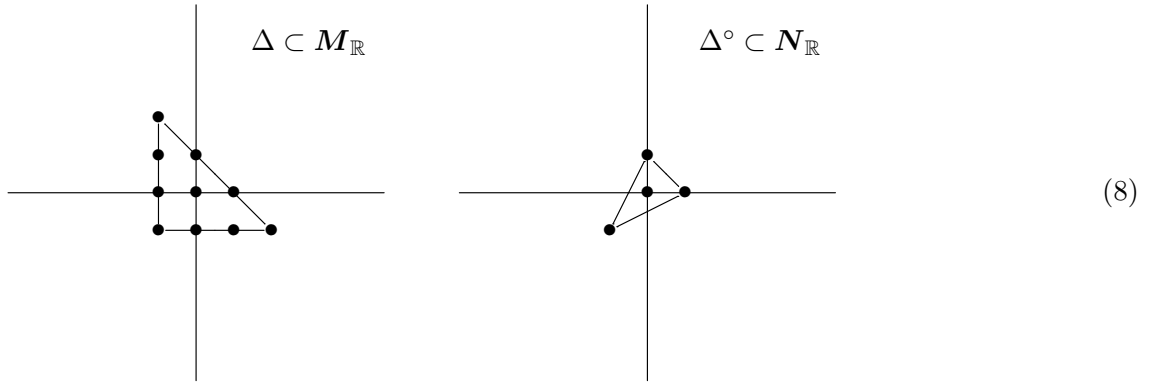
\begin{gather}
\label{eq:P2}
\begin{xy} <1.0mm,0mm>:
(-30,0)*\xybox{
  (0,0)*{} ="0", (25,0)*{}="pX", (-25,0)*{}="mX",
  (0,25)*{} = "pY", (0,-25)*{} ="mY", (15,20)*{\Delta \subset \bM_\R},
 (-5, 10)*{\bullet}="v1", (-5, 5)*{\bullet}, (-5,0)*{\bullet}, (-5,-5)*{\bullet}="v2",   ( 0,   5)*{\bullet}, (0, 0)*{\bullet}, (0,-5)*{\bullet}, ( 5,  0)*{\bullet}, (5, -5)*{\bullet}, (10,-5)*{\bullet}="v3",\ar@{-}|{} "mX"; "pX"
\ar@{-}|{} "mY"; "pY"
\ar@{-}|{} "v1"; "v2"
\ar@{-}|{} "v2"; "v3"
\ar@{-}|{}"v3"; "v1"
} ,
(30,0)*\xybox{
  (0,0)*{} ="0", (25,0)*{}="pX", (-25,0)*{}="mX",
  (0,25)*{} = "pY", (0,-25)*{} ="mY", (15,20)*{\Delta^\circ \subset \bN_\R},
  (0,5)*{\bullet}="v1", (5,0)*{\bullet}="v2" , (0,0)*{\bullet} , (-5,-5)*{\bullet}="v3",
\ar@{-}|{} "mX"; "pX"
\ar@{-}|{} "mY"; "pY"
\ar@{-}|{} "v1"; "v2"
\ar@{-}|{} "v2"; "v3"
\ar@{-}|{}"v3"; "v1"
}
\end{xy}
\end{gather}
The polytope $\Delta$ may be thought of as the Newton polytope for a
hypersurface $\{P = 0\} \in (\mathbb{C}^\ast)^d$ which has a natural
compactif\/ication to a subvariety $X = \overline{ \{P = 0\}}$ in a toric
variety $V$ with fan $\Sigma_V \subset \bN_{\R}$, given by taking cones over
faces of $\Delta^\circ$.  When $\Delta$ is ref\/lexive then $X \subset V$ is a~Calabi--Yau hypersurface with ``suitably mild'' Gorenstein singularities;  for instance, when $d=4$ the generic hypersurface is smooth.

This construction has two notable virtues.  First, it gives a combinatorial condition (i.e.\ ref\/lexivity) for constructing many examples of smooth Calabi--Yau three-folds; this was used to great ef\/fect in~\cite{Kreuzer:2000xy} to produce the largest know set of such manifolds.  Second, since the ref\/lexive polytopes occur in pairs, there is a simple conjecture to produce the mirror of $X$:  we just construct the dual hypersurface $X^\circ \subset V^\circ$ by interpreting $\Delta^\circ$ as the Newton polytope for the hypersuface and $\Delta$ as def\/ining the toric variety $V^\circ$.  Mirror symmetry predicts the equality of Hodge numbers $h^{1,1}(X) = h^{1,2}(X^\circ)$ and $h^{1,2} (X) = h^{1,1}(X^\circ)$, and the conjectured pairing passes this important test.

The equality of the Hodge numbers is a f\/irst step in matching the moduli spaces of deformations of the Calabi--Yau manifolds and the corresponding (2,2) conformal f\/ield theories.  Recall that at a generic point, the moduli space associated to $X$ is a product of two special K\"ahler manifolds, the complex structure moduli space $\cM_{cx}(X)$ and the complexif\/ied K\"ahler moduli space~$\cM_{cK}(X)$.  The tangent spaces of these manifolds are canonically identif\/ied with inf\/initesimal deformations:
\[
 T_{\cM_{cx}} \simeq H^1(X,T_X), \qquad T_{\cM_{cK}} \simeq H^1(X,T^\ast_X).
 \]
The special K\"ahler metrics are determined from two holomorphic prepotentials $\cF_{cx}(X)$ and $\cG_{cK}(X)$, and the mirror map $ \cM_{cK}(X) \to \cM_{cx}(X^\circ)$ is an isomorphism of the moduli spaces as special K\"ahler manifolds.  The resulting relation $\cG_{cK}(X) = \mu^\ast   \cF_{cx}(X^\circ)$ leads to the celebrated relations between Gromov--Witten invariants of $X$ encoded in $\cG_{cK}(X)$, and the variations of Hodge structure on the mirror $X^\circ$ encoded by $\cF_{cx}(X^\circ)$.

\subsubsection*{The monomial-divisor mirror map}

The equality of Hodge numbers and the mirror isomorphism of the moduli spaces receives an important ref\/inement in the context of Batyrev mirror pairs $X\subset V$ and $X^\circ \subset V^\circ$.

A simple set of complex structure deformations of $X$ is obtained by considering variations of the def\/ining hypersurface modulo automorphisms of the ambient toric variety.  The resulting subspace of ``polynomial'' deformations has complex dimension
\[
%\label{eq:h12}
h^{1,2}_{\text{poly}} = \ell(\Delta) - d -1- \sum_{\vphi} \ell^\ast(\vphi) ,
\]
where $\ell$ counts the number of lattice points contained in a closed subset,
$\vphi$ is a facet of~$\Delta$, and~$\ell^\ast$ counts the number of lattice
points in the relative interior of the indicated closed subset.
We can understand this number as follows:  $\ell(\Delta)$ is the number of
monomials $\mu$ in the def\/ining polynomial~$P$; the group of connected
automorphisms of $V$ contains the $(\mathbb{C}^\ast)^d$ action which can be
used to rescale $d$ of the coef\/f\/icients to, say, $1$, and of course an overall
rescaling of $P$ does not af\/fect $X$; f\/inally, there are additional automorphisms of $V$ that can be used to set the coef\/f\/icients of monomials $\mu \in \relint(\phi)$ to zero.

Similarly, there is a simple way to obtain a subset of complexif\/ied K\"ahler deformations by taking the classes dual to the ``toric divisors'' on $X$, i.e.\ those obtained by pulling back toric divisors from the ambient space $V$.  These are counted by
\begin{gather}\label{eq:h11}
h^{1,1}_{\text{toric}} = \ell(\Delta^\circ) - d -1-\sum_{\vphi^\circ} \ell^\ast(\vphi).
\end{gather}
This count also has a simple interpretation:  the f\/irst three terms count the toric divisors on $V$ if we take the one-dimensional cones to be all lattice points in $\Delta \backslash\{0\}$; we subtract the last term since a toric divisor in $\relint{\vphi^\circ}$ does not intersect a generic hypersurface.

{\sloppy In general, in addition to these ``simple''  deformations, a variety $X \subset V$ has both non-polynomial complex structure deformations, and non-toric deformations of complexif\/ied K\"ahler structure.  Remarkably, however, $h^{1,1}_{\text{toric}}(X) = h^{1,2}_{\text{poly}}(X^\circ)$!  It is then natural to conjecture a~restricted isomorphism
\[
\cM_{cK}^{\text{toric}}(X) = \cM_{cx}^{\text{poly}}(X^\circ).
\]
As we will review in more detail below, this leads to the
Monomial-Divisor Mirror Map (MDMM)~\cite{Aspinwall:1993rj}.

}

\subsubsection*{The (2,2) gauged linear sigma model}

The MDMM isomorphism is particularly natural in the context of the Gauged Linear Sigma Model (GLSM) construction~\cite{Morrison:1994fr, Witten:1993yc}.  A (2,2) GSLM is a two-dimensional Abelian gauge theory with (2,2) supersymmetry that can be constructed from the data of $X \subset V$.  The theory naturally incorporates the two sets of deformations into two holomorphic superpotentials~$W$ and~$\Wt$; the former encodes the complex coef\/f\/icients of~$P$, while the latter encodes the complexif\/ied K\"ahler deformations of the ambient space $V$.

While the GLSM is not a conformal f\/ield theory, it is believed to reduce to an appropriate (2,2) conformal f\/ield theory at low energies.  Of course not all naive parameters contained in~$W$ and~$\Wt$ correspond to genuine deformations of the conformal f\/ield theory.  In fact, holomorphic f\/ield redef\/initions that do not af\/fect the low energy theory can be used to reduce the deformations in~$W$ to the $h^{1,2}_{\text{poly}}(M)$ deformations described above~\cite{Kreuzer:2010ph}.  These GLSM parameters yield algebraic (as opposed to special K\"ahler) coordinates on the moduli space; in terms of these algebraic coordinates the MDMM takes a canonical form.

Furthermore, the GLSM admits the A- and B-topological twists which
correspond to the A- and B-model topological subsectors of the conformal f\/ield
theory~\cite{edtft}.  By using the methods of toric residues, combined with
summing the instantons in GLSMs it is possible to show that the MDMM indeed
exchanges the observables of the A- and
B-models~\cite{Batyrev:2002em,Morrison:1994fr,Szenes:2004mv}.  This is one of
the most convincing and important tests of the mirror correspondence\footnote{These results have also been generalized to complete intersections in toric varieties~\cite{Borisov:2005hs,Karu:2005tr}.}.

\subsubsection*{Correlators and the discriminant locus for the quintic}

To illustrate some salient features, we review the GLSM presentation~\cite{Morrison:1994fr} of the original mirror computation for the quintic in algebraic coordinates~\cite{Candelas:1990qd}.  In this case $h^{1,1}(X) = 1$, and the A-model depends on one complexif\/ied K\"ahler parameter $q$.  The GLSM contains an operator $\sigma$ that corresponds to the inf\/initesimal deformation $H^1(X,T^\ast_X)$, and the single $A$-model correlator is given by
\begin{gather}
\label{eq:Acorr}
\langle \sigma^3 \rangle_{A} : \ \Sym^3 H^1(X,T^\ast_X) \to \mathbb{C}, \qquad
\langle \sigma^3 \rangle_{A} = \frac{5}{1+ 5^5 q}.
\end{gather}
The mirror quintic is def\/ined is a hypersuface in $\P^4/{\Z_5^3}$ with
\[
 P^\circ = Z_0^5 + Z_1^5 + Z_2^5 + Z_3^5+Z_4^5  - 5 \psi Z_0Z_1Z_2Z_3Z_4.
\]
We used the $(\mathbb{C}^\ast)^5$ rescalings to bring $P^\circ$ into a
canonical form; the parameter $\psi$ is a coordinate on (a f\/ive-fold cover)
of the complex structure moduli space $\cM_{cx}(X^\circ)$, and the mirror GLSM contains an operator $\mu$ that corresponds to the inf\/initesimal deformation $H^1(X^\circ, T_{X^\circ})$.  Up to a~$\psi$-independent constant, the $B$-model correlator is given by
\begin{gather}
\label{eq:Bcorr}
\langle \mu^3 \rangle_{B} : \ \Sym^3 H^1(X^\circ, T_{X^\circ}) \to \mathbb{C},
\qquad
\langle \mu^3\rangle_{B} = \frac{5^{4}\psi^2}{1 - \psi^5}.
\end{gather}
The monomial-divisor mirror map in this case reads $q = (-5\psi)^{-5}$,
and it exchanges the two correlators provided we identify the deformations as
$\sigma \sim q \ff{\partial}{\partial q}$ and
$\mu \sim \psi \ff{\partial}{\partial \psi}$.\footnote{This exchange is again
up to an overall independent constant; the ambiguity can be resolved by
working with special coordinates and physical Yukawa couplings~\cite{Candelas:1990qd}.}  These transformations are a consequence of topological descent in the A and B models~\cite{Dijkgraaf:1990qw}.  Note also that the correlators diverge at $q = -5^{-5}$ or, equivalently, at $\psi^5 = 1$.  The B-model divergence at $\psi^5=1$ is easy to see:  this is exactly the discriminant locus where the hypersurface $P^\circ=0$ is singular.  The A-model divergence at $q=-5^{-5}$ is due to a divergent sum over the GLSM instantons. In more general theories the discriminant locus has many components, but one can show that the MDMM exchanges the discriminant loci of a pair of mirror theories.

\subsubsection*{Redundant deformations and plain polytopes}

The presentation of the toric and polynomial deformations is complicated by the ``redundant'' monomials corresponding to points in $\relint(\vphi)$ and the ``redundant'' divisors corresponding to points in $\relint(\vphi^\circ)$.  In fact, in the context of topological f\/ield theory and GLSM computations, while it is relatively easy to see that f\/ield redef\/initions can be used to eliminate the redundant monomials, it is not so simple to understand the decoupling of deformations corresponding to the redundant divisors.\footnote{It is trivial at the level of classical geometry; however, showing it at the level of GLSM gauge instantons is more involved~\cite{Kreuzer:2010ph}.}  Thus, computations in the (2,2) setting become simpler if these redundant monomials and divisors are absent.  To quantify this absence, we say a polytope $\Delta$ is plain if none of its facets contains an interior lattice point.  Thus, in~(\ref{eq:P2}) $\Delta^\circ$ is plain, while $\Delta$ is not plain.  A ref\/lexive polytope $\Delta$ is ref\/lexively plain if both it and its dual are plain.  In two dimensions there is a single self-dual ref\/lexively plain polytope:
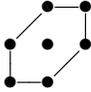
\begin{gather*}
\begin{xy} <1.0mm,0mm>:
  %(15,20)*{\Delta \subset M_\R},
(0,0)*{\bullet}, (5, 5)*{\bullet}="v1",  ( 0, 5)*{\bullet}="v2",(-5,0)*{\bullet}="v3", (-5,-5)*{\bullet}="v4",
 (0,-5)*{\bullet}="v5", ( 5,  0)*{\bullet}="v6",
%\ar@{-}|{} "mX"; "pX"
%\ar@{-}|{} "mY"; "pY"
%\ar@{-}|{} "v1"; "v2"
%\ar@{-}|{} "v2"; "v3"
\ar@{-}|{}"v1"; "v2"
\ar@{-}|{}"v2"; "v3"
\ar@{-}|{}"v3"; "v4"
\ar@{-}|{}"v4"; "v5"
\ar@{-}|{}"v5"; "v6"
\ar@{-}|{}"v6"; "v1"
\end{xy}
\end{gather*}
The $473,800,776$ ref\/lexive polytopes in $d=4$ have $6,677,743$ ref\/lexively plain non-self-dual pairs and $5,518$ self-dual ref\/lexively plain polytopes~\cite{Kreuzer:2010ph}.  A simple example of a $d=4$ ref\/lexively plain pair has vertices
\begin{gather}
\label{eq:quinticfriend}
\Delta    : \begin{pmatrix} 1 & 0 & 2 & 3 & -6 \\ 0 & 1 & 4 & 3 & -8 \\ 0 & 0 & 5 & 0 & -5 \\ 0 & 0 & 0 & 5 & -5 \end{pmatrix},\qquad
\Delta^\circ         : \begin{pmatrix} -1 & -1 & 1 & 1 \\ -1 & -1 & 1 & 2 \\  -1 & -1 &2 & 1 \\ -1 & 4 & -3 & -2 \\ 4 & -1 & -1 &-2\end{pmatrix}.
\end{gather}
$\Delta$ has a total of $26$ lattice points, while $\Delta^\circ$ has no additional non-zero lattice points.  The corresponding three-fold is the $\Z_5$ quotient of the quintic in $\C\P^4$ with
\[
h^{1,1}(X) = h^{1,1}_{\text{toric}}(X) = 1\qquad \text{and}\qquad
 h^{1,2}(X) = h^{1,2}_{\text{poly}}(X) = 21.
\]

\subsection{(0,2) Gauged linear sigma models and a mirror map}

Having reviewed some basic structure of (2,2) mirror symmetry, we are now ready to discuss the (0,2) deformations and a possible mirror map.  The f\/irst naive guess is to start with the pair $(X,T_X)$ and $(X^\circ, T_{X^\circ})$ and try to match deformations of $T_X$ and $T_{X^\circ}$ as holomorphic bundles.  This runs into two problems.  The f\/irst is familiar from classical algebraic geometry.  Unlike the f\/irst order deformations $H^1(X,T_X)$ and $H^1(X,T_X^\ast)$, which can be integrated to f\/inite deformations of complex and K\"ahler structure, respectively, the inf\/initesimal deformations of the bundle, characterized by $H^1(X, \End T_X)$ can have higher order obstructions.  The second issue has to do with quantum obstructions.  In general, a (0,2) supersymmetry-preserving deformation of a (2,2) theory need not preserve conformal invariance, so that turning on a classically unobstructed deformation in $H^1(X,\End T_X)$ can ruin the structure of the conformal f\/ield theory.  For instance, in the case of the famous ``Z-manifold'', i.e.\ the resolution of $T^6/\Z_3$ with $h^{1,1} = 36$ and $h^{1,2} = 0$ it is known that $h^1(X,\End T_X) = 208$, but $108$ of these are obstructed at f\/irst order by worldsheet instanton ef\/fects~\cite{Aspinwall:2011us}.

\subsubsection*{(0,2) GLSM deformations}

Fortunately, the GLSM construction helps with both the classical and quantum obstructions.  The original observation, going back to~\cite{Witten:1993yc}, is that the (2,2) GLSM Lagrangian, viewed as a~(0,2) theory, has holomorphic (0,2) deformation parameters encoded in the following complex of sheaves on~$X$
\begin{gather}
\label{eq:Defs}
\xymatrix{ 0\ar[r] &\cO^{r}|_X \ar[r]^-{E} &\oplus_\rho\cO(D_\rho)|_{X}  \ar[r]^-{J} & \cO(\sum_\rho D_\rho)|_{X}\ar[r] & 0 }.
\end{gather}
Here the ambient toric variety $V$ is presented as a holomorphic quotient $\{\C^n~\backslash~F\}/(\C^\ast)^r$, and the $D_\rho$ are the toric divisors on $V$.  The cohomology of this complex, $\cE = \ker J / \im E$, def\/ines a rank $d-1$ holomorphic bundle over~$X$ that is a deformation of the tangent bundle $T_X$.  The maps $E$ and $J$ have a simple form on the (2,2) (i.e.~$\cE = T_X$) locus.  The $J$ are the dif\/ferentials of the def\/ining equation, $dP$, while $E$ is the familiar map from the Euler sequence for the tangent bundle of the toric variety:
\begin{gather}\label{euler}
\xymatrix{ 0\ar[r] &\cO^{r} \ar[r]^-{E} &\oplus_\rho\cO(D_\rho) \ar[r] & T_V\ar[r] & 0 }.
\end{gather}
Just as the complex coef\/f\/icients of the def\/ining equation over-parametrize the space of polynomial complex structure deformations, so do the coef\/f\/icients in $E$ and $J$ over-parametrize the space of ``monadic'' bundle deformations because of various
automorphisms of the complex~\cite{Kreuzer:2010ph}.  However, these can be taken into account and a combinatorial formula, akin to~(\ref{eq:h11}), gives the total number of monadic deformations.

For instance, for the Fermat quintic in $\C\P^4$ we have
\[
r=1, \qquad E = (Z_0,Z_1,Z_2,Z_3,Z_4)^T, \qquad J = \big(Z_0^4,Z_1^4,Z_2^4,Z_3^4,Z_4^4\big).
\]
Since $\sum_\rho E^\rho J_\rho = 5 P$ this is indeed a complex of sheaves on~$X$.  By identifying the automorphisms of the complex, we f\/ind that the monadic deformations yield a $224$-dimensional space of deformations.  Since that is exactly $h^1(X,\End T_X)$ in this case, we see that all inf\/intesimal deformations of the quintic's tangent bundle are unobstructed.

Given this complex, we have a simple way of obtaining a family of sheaves $\cE$ over $X$ by deforming the def\/ining maps $E$ and $J$ in~(\ref{eq:Defs}) while preserving $J \circ E|_{X} = 0$.  Moreover, there exists a general set of arguments that holomorphic deformations of a GLSM are protected from worldsheet instanton destabilization~\cite{Basu:2003bq,Beasley:2003fx, Silverstein:1995re}.  In particular, the work of~\cite{Beasley:2003fx} gives a method that can be used to show that toric worldsheet instantons do not destabilize a (0,2) GLSM.  It should be stressed that this result has not yet been formulated  as a general vanishing theorem, and in principle it must be checked example by example; however, the structure of the argument suggests that a general formulation may be possible.  At any rate, for many models the argument is indeed suf\/f\/icient.

In general the holomorphic parameters of the (0,2) GLSM cannot describe all of the bundle deformations; this should not surprise us, since even on the (2,2) locus the GLSM only captures the toric K\"ahler and polynomial complex structure deformations.  The ``monadic'' bundle deformations encoded in~(\ref{eq:Defs}) do, however, of\/fer a simple parametrization of a subset of unobstructed (0,2) deformations.  They have been the focus (and the secret of success) for much recent work in (0,2) theories.

Note that the bundle complex depends on the complex structure of~$X$.  This illustrates an  important point:  in general there is no invariant way to split (0,2) deformations into the sort of canonical form familiar from the (2,2) context.  While it is known that the full (0,2) moduli space must be K\"ahler~\cite{Periwal:1989mx}, it need not have any canonical product form akin to that familiar from the (2,2) context, nor does it need to admit a special K\"ahler metric.

The quintic once again provides a simple example of the general structure.  As we mentioned above, $h^1(X,\End T_X) =224$ for the quintic, and all of these deformations
are classically unobstructed.  Moreover, all of them can be represented by deforming the $E$ and $J$ maps, and the method of~\cite{Beasley:2003fx} immediately shows that there are no quantum obstructions either.  Thus the full (0,2) quintic moduli space has dimension
\[
h^1(X,T^\ast_X) + h^1(X,T_X)+ h^1(X,\End T_X) = 326.
\]
In the context of mirror symmetry it is natural to ask how to represent these deformations in the GLSM for the mirror quintic.  Are the $224$ (0,2) deformations representable by deforming the mirror complex?  Unfortunately, the answer is no.  The mirror of~(\ref{eq:Defs}) only yields $164$ deformations, so that $60$ additional deformations remain unaccounted~\cite{Kreuzer:2010ph}.

Of course this is not a failure of mirror symmetry but merely a dif\/f\/iculty in presenting the deformations in a useful fashion.  Although to our knowledge the computation of
$h^1(X^\circ, \End T_{X^\circ})$ has not been carried out, the conformal f\/ield theory mirror isomorphism on the (2,2) locus implies that this should be $224$ as well; moreover, all of these should be unobstructed, both at the classical and quantum levels.  However, studying these deformations and any possible mirror map abstractly appears to be rather dif\/f\/icult.

It was observed in~\cite{Kreuzer:2010ph} that this mismatch of (0,2) GLSM deformations for Calabi--Yau hypersurfaces in toric varieties is a rather general feature among the ref\/lexive polytope pairs.  This should be contrasted with the situation on the (2,2) locus, where the matching of GLSM parameters to those of the mirror GLSM corresponds to matching $\cM^{\text{toric}}_{cK} (X)$ to $\cM^{\text{poly}}_{cx}(X^\circ)$.  However, it was also observed that precisely for the case of ref\/lexively plain polytope pairs the number of  GLSM (0,2) deformations matched as well.  For instance, in the case of the hypersurface in~(\ref{eq:quinticfriend}) there are $39$ (0,2) ``monadic'' deformations, giving a deformation space of dimension $66$ on both sides of the mirror.

\subsubsection*{A (0,2) mirror map}

A pair of GLSMs associated to a ref\/lexively plain polytope and its dual is a natural candidate for f\/inding an explicit mirror map for the full GLSM moduli space.  The inspiration for this comes directly from the monomial-divisor mirror map~\cite{Aspinwall:1993rj}, where the equality  $h^{1,1}_{\text{toric}}(X) = h^{1,2}_{\text{poly}}(X^\circ)$, combined with the action of the toric automorphism group $\Aut(V)$ as encoded by the combinatorial structure of $\Delta$, yielded a simple Ansatz for the map.  The analogue (0,2) analysis was performed in~\cite{Melnikov:2010sa}, yielding a concrete proposal for a (0,2) mirror map.  We will now review that result, but to present it we will need to introduce a little more combinatorial structure.

Let $\Delta,\Delta^\circ$ be a $d$-dimensional ref\/lexively plain pair as above.  Let $\Sigma_V$ be a maximal projective subdivision, so that one-dimensional cones $\rho \in\Sigma_V(1)$ are in one-to-one correspondence with the $n$ non-zero lattice points in $\Delta^\circ$.  Let $\C[Z_1, \ldots, Z_n]$ denote the homogeneous Cox ring for $V$, with the $(\C^\ast)^r$ action given by\footnote{In general $V = \{\C^n\backslash F\}/ G$, where $G = (\C^\ast)^r \times H$ and $H$ is a discrete Abelian group; as we will focus on automorphisms connected to the identity, $H$ will not play an important role in what follows.}
\begin{gather}
\label{eq:torus}
(t_1,\ldots,t_r) \cdot Z_\rho \to Z_\rho \times \prod_{a=1}^r t_a^{Q^a_\rho} .
\end{gather}
In terms of these homogeneous coordinates, the hypersurface takes the form
\[
P = \sum_{m\in \Delta\cap \bM} \alpha_m \mon_m, \qquad\text{with}  \quad \mon_m \equiv \prod_\rho Z_\rho^{\langle m,\rho\rangle +1},
\]
where $\alpha_m$ are the complex coef\/f\/icients of the monomials $\mon_m$.  To simplify some manipulations we will take $\alpha_m \neq 0$.
It is useful to introduce the rank $d$ matrix
\[
\pi_{m\rho} \equiv \langle m,\rho \rangle \qquad\text{for}\quad
m \in \{\Delta\cap \bM \backslash 0\}.
\]
The charges $Q^a_\rho$ form an integral basis for the $r$-dimensional kernel of $\pi$; similarly, the integral basis $\Qh^{\ah}_{m}$ for the cokernel of $\pi$ def\/ine the $(\C^\ast)^{r^\circ}$ action in the mirror toric variety
\[
V^\circ = \{ C^{n^\circ} \backslash F^\circ\}/ (\C^\ast)^{r^\circ}.
\]
Since we will assume that $\Sigma_{V^\circ}$ is also a maximal projective subdivision, $n^\circ = \ell(\Delta) -1$ and $r^\circ = \ell(\Delta) -1-d$ in the ref\/lexively plain case.

The maps of~(\ref{eq:Defs}) are
given by
\[
E^{a\rho} = e^{a\rho} Z_\rho, \qquad Z_\rho J_\rho   =
\sum_{m\in \Delta\cap \bM} j_{m\rho}\mon_m
\]
with $j_{m\rho} = 0$ whenever $\langle m,\rho \rangle  = -1$.
On the (2,2) locus we have
\[
E^{a\rho} = Q^a_\rho \qquad\text{and}\qquad j_{m\rho}   =
(\langle m,\rho\rangle+1) \alpha_m.
\]
In order for (\ref{eq:Defs}) to be a complex we require,
\begin{gather}
\label{eq:constraint}
\sum_\rho J_\rho(Z) E^{a\rho}(Z) + \delta^a P(Z) = 0,
\end{gather}
for some complex coef\/f\/icients~$\delta^a$.  Note that since $P$ transforms equivariantly under the torus action~(\ref{eq:torus}), the constraint holds automatically on the (2,2) locus with $\delta^a = -\sum_\rho Q^a_\rho$.

In addition to these parameters, the (0,2) GLSM also depends on the~$q_a$~-- the $r$ complexif\/ied K\"ahler parameters.
As we described in the discussion of (2,2) theories, these parameters occur in orbits corresponding to holomorphic f\/ield redef\/initions of the GLSM; with the exception of the action on the $q_a$ (a feature that only shows up in (0,2) theories), the action on the parameters arises via automorphisms of the toric variety and induced automorphisms on \eqref{eq:Defs}.  By examining the action of redef\/initions on the parameter space, we can construct a natural set of invariants from the~$q_a$, $\alpha_m$, $j_{m\rho}$, $e^{a\rho}$ and $\delta^a$ that satisfy the constraint~(\ref{eq:constraint}):
\begin{gather*}
\kappa_a \equiv q_a \prod_\rho \left( \frac{j_{0\rho}}{\alpha_0}\right)^{Q^a_\rho},\qquad
\kappah_{\ah} \equiv \prod_{m \in \{ \Delta\cap \bM\backslash0\}} \left(\frac{\alpha_m}{\alpha_0}\right)^{\Qh^{\ah}_m}, \nonumber\\
b_{m\rho} \equiv \frac{\alpha_0 j_{m\rho}}{\alpha_m j_{0\rho}} -1, \qquad \text{for} \quad m\neq 0.%\label{eq:coords}
\end{gather*}
In order for this data to def\/ine a non-singular theory, $b_{m\rho}$ must have rank exactly $d$.

On the (2,2) locus $b_{m\rho} = \pi_{m\rho}$, $\kappa_a = q_a$, and the $\kappah_{\ah}$ are the usual invariant algebraic coordinates on $\cM^{\text{poly}}_{cx} (X)$.  The MDMM then takes a simple form:  we repeat the construction of invariants for $X^\circ = \{P^\circ = 0\} \subset V^\circ$, which leads to coordinates $\kappa^\circ_{\ah}$ and $\kappah^\circ_{a}$.  The map is then
\begin{gather*}
%\label{eq:MDMM}
\kappa_a  = \kappah^\circ_a, \qquad \kappah_{\ah} = \kappa^\circ_{\ah}.
\end{gather*}
The proposed (0,2) extension is simple:  the rank $d$ matrix $b^\circ_{\rho m}$ of the mirror theory is just the transpose of~$b$.

\subsubsection*{Testing the proposal}

How can we test the proposed (0,2) mirror map?  The most direct way would be
to study the analogue of the A- and B-model correlators as
in~(\ref{eq:Acorr}) and~(\ref{eq:Bcorr}).  As discussed in
Section~\ref{quantumsheaf}, the analogous correlators
certainly
exist, see e.g.~\cite{Adams:2003d,Adams:2005tc,Donagi:2011uz,Donagi:2011,Katz:2004nn,McOrist:2008ji,s2}
in the A/2 and B/2 theories;
when $T_X$ is deformed to $\cE$, they correspond to multilinear maps
\begin{gather*}
\text{A/2}: \ \Sym^{d-1} H^1(X,\cE^\ast) \to \C, \qquad
\text{B/2}: \ \Sym^{d-1} H^1(X, \cE) \to \C.
\end{gather*}
However, the computations remain dif\/f\/icult, and general techniques remain
fairly undeveloped on the B/2-model side.

Although the full correlators remain out of reach, it turns out to be possible
to identify an important component of the singular locus~-- the subvariety in
the moduli space where the theory becomes singular.
How do these singularities arise?  In the A/2 model they come  from
diverging sums over the GLSM gauge instantons, while in the B/2 model they
have a classical manifestation:  as we change parameters in the $J$ maps the
sheaf $\cE$ def\/ined by~\eqref{eq:Defs} can fail to be a vector bundle by developing singularities.

As in the case of singularities in complex structure moduli space~\cite{Kapranov:1991}, where $P$ fails to be transverse in $V$, the resulting discriminant locus is in general reducible.  The principal component of the discriminant locus corresponds to degenerations that occur in $X \cap (\C^\ast)^d \subset V$, and it is characterized in the following fashion~\cite{Melnikov:2010sa}.  Let $\gammah^m_{\ah}$ parametrize the cokernel of $b_{m\rho}$ and let $\deltah^m \equiv -\sum_{\ah} \gammah^m_{\ah}$.  The bundle $\cE$ is then singular at some point in $X \cap (\C^\ast)^d \subset V$ if and only if there exists a~vector $\sigmah \in \C^{r^\circ}$ satisfying
\[
\prod_{m \in \{ \Delta\cap \bM \backslash 0 \}} \left[ \frac{\sigmah\cdot \gammah^m}{\sigmah \cdot \deltah} \right]^{\Qh^{\ah}_m} = \kappah_{\ah},
\]
where $\sigmah \cdot \gammah^m \equiv \sum_{\ah} \sigmah^{\ah} \gammah^m_{\ah}$.  For the A/2 model the principal component of the discriminant locus was computed in~\cite{McOrist:2008ji}.  The result, recast in the invariant coordinates, is that there must exist a~vector $\sigma \in \C^r$ satisfying
\[
\prod_{\rho} \left[ \frac{\sigma\cdot \gamma^\rho}{\sigma \cdot \delta} \right]^{Q^a_\rho} = \kappa_{a},
\]
where $\gamma^\rho_a$ form a basis for the kernel of $b_{m\rho}$
and $\delta^\rho \equiv -\sum_a \gamma^\rho_a$.   It is now evident that under
the proposed mirror map the principal component of the discriminant of the
A/2 (B/2) model will be mapped to the principal component of the discriminant
of the mirror $\text{B}^\circ/2$ ($\text{A}^\circ/2$) model.  This non-trivial check constitutes the best to-date evidence for the (0,2) mirror map in ref\/lexively plain examples.

\section{Worldsheet duality}
\label{worldsheetduality}

In the previous section, we decribed a proposal for a (0,2) monomial-divisor
mirror map, at least for ref\/lexively plain polytopes and bundles that
are deformations of the tangent bundle.  However, we did not discuss
(0,2) mirrors for non-Calabi--Yau spaces or models without a (2,2) locus, nor did we give any attempt
at a physical derivation.

For (2,2) mirror symmetry, there have been very interesting attempts to derive mirror
sym\-met\-ry from a GLSM construction in~\cite{Morrison:1995yh}, and from worldsheet duality
in~\cite{Hori:2000kt}. The latter approach provides a nice construction of a mirror description of a GLSM with a toric target space, inclu\-ding non-compact toric Calabi--Yau spaces. It is one of the major open questions of both (2,2) and (0,2) mirror symmetry to f\/ind a derivation that applies to compact Calabi--Yau spaces.

Reducing the worldsheet supersymmetry from (2,2) to (0,2) greatly
enriches both the physical space of theories and their associated mathematical
structures, but at the expense of reduced control over the quantum dynamics
of these models. Fortunately, (0,2) supersymmetry still provides suf\/f\/icient
control that we can def\/ine and compute rings of observables.
These rings def\/ine quantum sheaf cohomology discussed in Section~\ref{quantumsheaf}.
In this section, we will explain how to construct mirror pairs
of (0,2) models using worldsheet duality~\cite{Adams:2003d},
generalizing the approach used in the (2,2) setting by
\cite{Hori:2000kt}.

%The approach can be used for $c_1\geq 0$ toric varieties
%This will not only give a physical
%derivation, but also non-Calabi--Yau mirrors, as well as make predictions
%for quantum sheaf cohomology, which will be reviewed in the next section.

\subsection{The basic idea}

%Linear sigma models are gauge theories coupled to charged matter.
 We will consider (0,2) gauged linear sigma models  discussed in Section~\ref{deformtangent}. We again restrict to Abelian gauge groups, but we will allow bundles which are not necessarily deformations of the tangent bundle of the target space toric variety $V$. Indeed, even the rank of $\cE$ can dif\/fer from the rank of $ T_V$. Since the construction is a worldsheet duality, we will need to introduce some notions from superspace to describe the worldsheet theories.
Our conventions can be found in Appendix~\ref{superfieldconventions}. The basic ingredients of such a theory are a collection of $U(1)^k$ gauge-f\/ields and coupled charged matter. The gauge-f\/ields reside in fermionic superf\/ields $\Upsilon^a$ with $a=1,\ldots, k$, while the matter  resides in chiral multiplets $\Phi^i$ with charges $Q^a_i$.
 Under a gauge transformation with parameters $\Lambda^a$,
\begin{gather*} %\label{gaugeaction}
\Phi^i \rightarrow e^{i \Lambda^a Q^a_i} \Phi^i.
\end{gather*}
In the most commonly studied (0,2) models, this data is suf\/f\/icient to determine the target space geometry which is the toric variety $V$.

The basic idea of Abelian duality is to implement T-duality along a $U(1)$ isometry direction of the target space. We can see how this duality works in a simple case. Consider a free theory consisting of a circle-valued scalar f\/ield $\phi$ with period $2\pi$. Instead of writing the action in the most straightforward way,  introduce a Lagrange multiplier $1$-form $A$ and consider the two-dimensional action:
\[ %\label{modaction}
S = {1\over 4\pi R^2} \int  A\wedge *A
      -{i\over 2\pi}     \int  \phi \,  dA.
\]
Integrating out $A$ in a quadratic theory like this amounts to solving the classical $A$ equation of motion,
\[
 A=-i R^2 * d\phi,
\]
which gives the free action for a scalar f\/ield on a circle of size $R$:
\[ %\label{trivial}
S = {R^2 \over 4\pi} \int  (\partial\phi)^2 .
\]
On the other hand, integrating out $\phi$ imposes the constraint  $dA=0$, which we can solve via $A = d{\widetilde \phi}$ with ${\widetilde \phi}$ periodic with period $2\pi$.   The dual action is therefore,
\[
S = {1\over 4\pi R^2} \int   (\partial\widetilde\phi)^2,
\]
which describes a scalar f\/ield on a circle of size $1/R$. Note that there is no local relation between~$\phi$ and~$\widetilde\phi$. The map between the two descriptions is non-local and involves an exchange of momentum and winding modes.

We will slightly generalize this procedure and separately dualize the $U(1)$ action acting on the phase of each chiral superf\/ield~$\Phi^i$.   The $U(1)$ action acting on the phase is not free. We will also gauge $k$ combinations of these $U(1)$ actions. The existence of a f\/ixed point for each~$U(1)$ is ref\/lected in the dual description in two ways: there is a non-perturbative term in the superpotential as well as a non-trivial dilaton f\/ield. This dualization procedure is a kind of
worldsheet analogue of the SYZ proposal for constructing mirror Calabi--Yau spaces~\cite{syz}, except it can be applied to both conformal and massive models with toric target spaces~$V$.

Now let us introduce additional data determining the target space gauge bundle. Abelian duality is best understood in models without a tree-level superpotential so we set $J=0$.\footnote{To be more precise: duality can currently be implemented to some degree with $E\neq 0$ or $J\neq 0$, but not with both couplings non-vanishing. The case of a compact Calabi--Yau space with a stable holomorphic bundle requires both sets of couplings.} As noted above, implementing duality with $J \neq 0$ is perhaps the central question in understanding mirror symmetry  for compact Calabi--Yau spaces. In this setting, the gauge bundle is encoded in a choice of left-moving fermionic chiral superf\/ields $\G^A$ satisfying
\[
\bar{\mathfrak{D}}_+ \G^A = \sqrt{2} E^A(\Phi),
\]
with gauge charges $Q^a_A$. Freedom from worldsheet gauge anomalies imposes a quadratic relation on
the gauge charges,
\[
\sum_{i} Q^a_i Q^b_i - \sum_A Q^a_A Q^b_A=0,
\]
for each $(a,b)$. The additional condition,
\[
\sum_i Q^a_i=0,
\]
is required for conformal invariance at one-loop.
These conditions guarantee that the infrared non-linear theory satisf\/ies the
anomaly cancelation condition~(\ref{gs-condition}) stated in the introduction.

The holomorphic $E^A$ couplings def\/ine the maps which determine the holomorphic bundle
over the toric variety~$V$. For a (2,2) model, there is a single Fermi multiplet $\G$ for each
chiral multiplet $\Phi$. The associated bundle is the tangent bundle def\/ined by the Euler sequence~(\ref{euler}) if we choose
\[
 E^i= Q^a_i \Sigma^a \Phi^i
 \]
with each $\Sigma^a$ a neutral chiral multiplet.
For more general models with  $E^A = \Sigma^a {\widehat E}^{Aa}(\Phi_i)$,
the left-moving fermions def\/ine a monad bundle via the exact sequence
\begin{gather}\label{geneuler}
\xymatrix{ 0\ar[r] &\cO^{r} \ar[r]^-{{ E}} &\oplus_A \cO(D_A) \ar[r] & \cE  \ar[r] & 0}.
\end{gather}
Note that even these choices of $E^A$ are quite special from the perspective of a general (0,2)
model. From a (0,2) perspective, there is no reason for a
distinguished $\Sigma^a$ multiplet at all! The $\Sigma^a$ multiplet is really a
vestige of (2,2) supersymmetry and the structure of a (2,2) vector
multiplet, yet there is a rich class of theories of this type as we have already seen.

The starting data is therefore a collection of charged f\/ields $(\Phi^i, \G^A)$,
a choice of complexif\/ied K\"ahler parameters $t^a = ir^a + {\theta^a\over 2\pi}$,
and a choice of holomorphic bundle specif\/ied by the  $E^A$. There are also gauge supermultiplets, $\Upsilon$, and neutral chiral superf\/ields, $\Sigma$, which come along for the ride. The result of the Abelian dualization procedure is a theory expressed in terms of neutral f\/ields $(Y^i, F^A)$ where
\begin{gather*}
{\rm Im} (Y^i) \sim {\rm Im} (Y^i) + 2\pi, \qquad {\rm Re} (Y^i)\geq 0,
\end{gather*}
with $Y^i$ a chiral superf\/ield and $F^A$ a neutral Fermi superf\/ield.  The dual theory is a (0,2) Landau--Ginzburg theory for which holomorphic quantities are controlled by a fermionic superpotential:
\begin{gather*} %\label{superpotential}
{L_W} =    \left(\int d \theta^+ {W} + {\rm h.c.}\right).
\end{gather*}
We will focus on the structure of this superpotential;
the full Lagrangian can be found in~\cite{Adams:2003d}. Since $W$ is fermionic, it takes the form
\[
W = \Gamma \cdot J,
\]
where $\Gamma$ denotes a collection of Fermi superf\/ields and $J$ denotes a collection of functions of chiral superf\/ields. Supersymmetric vacua are found by solving $J=0$.

Fortunately, supersymmetry combined with $R$-symmetry is suf\/f\/icient to determine the general form of the non-perturbative superpotential in  models both with and without a (2,2) locus. The exact dual superpotential is given by,
\begin{gather} \label{exact}
{{W}} = \sum_a \left(
-\frac{i \Upsilon^a}{4}\left(\sum_i Q_i^a Y^i + it^a \right)
+ \frac{\Sigma^a}{\sqrt 2} \sum_A  Q^{a}_A F^A \right)
+ \mu \sum_{i A} \beta_{i A} F^A e^{-Y^i},
\end{gather}
where we have made  the mass scale $\mu$ explicit. The neutral $Y^i$ f\/ields, dual to the charged f\/ields $\Phi^i$, are axially coupled to the gauge-f\/ield. This f\/irst term of~(\ref{exact}) proportional to $\Upsilon^a Q^a_i Y^i$ corresponds to a dynamical theta-angle, which ref\/lects this axial coupling.
The last term
in~(\ref{exact}) is an instanton-induced coupling. The dependence on the
K\"ahler parameters $t^a$ is explicit. The parameters $\beta_{i A}$ are the
interesting quantities corresponding to the choice of
holomorphic gauge bundle.

Determining the map between the $\beta_{i A}$ and the original $E^A$
parameters is the most challenging step. The way the map has been understood so far is by working on the
Coulomb branch of both the original and dual models where $\Sigma^a$ has an
expectation value, rather than the Higgs branch where the $\Phi^i$ have expectation values.
In simple cases, matching the expressions for the ef\/fective potential for
${{W}}(\Sigma, \Upsilon)$ in both the original and dual descriptions determines the relation
between some of the bundle deformation parameters and the $\beta$-parameters.

\subsection{The duals of Hirzebruch surfaces}

As a new example, let us consider a linear model with target space a Hirzebruch
surface~${\mathbb F}_n$. The gauged linear sigma model exists for any~$n$; however, a non-linear sigma model with target~${\mathbb F}_n$ is only asymptotically free for $n<3$.
The gauge theory with~${\mathbb F}_n$
as a moduli space has a~$U(1) \times U(1)$ gauge group. We introduce four
chiral superf\/ields with charges $(1,0)$, $(0,1)$, $(1,0)$ and  $(n,1)$. We can leave the holomorphic bundle unspecif\/ied for the moment. It is specif\/ied by a choice of~$\G^A$ Fermi superf\/ields in the gauge theory.

The mirror Landau--Ginzburg theory is then described in terms of four $Y^i$ and a collection of neutral Fermi superf\/ields $F^A$. The mirror superpotential is of the form given in~(\ref{exact}). The ground state structure is determined as follows: f\/irst, we need to solve the two $\Upsilon$ constraints,
\begin{gather}\label{constraints}
Y^1 + Y^3+ nY^4 + it^1=0, \qquad Y^2 + Y^4 + it^2=0,
\end{gather}
which follow from setting $\left(\sum_i Q_i^a Y_i + it^a \right)=0$ with $a=1,2$. The solution space is two-di\-men\-sional, and we can choose a convenient integral basis spanned by the two vectors $(1,0,-1,0)$, $(0,-1,-n,1)$. In terms of this basis, we solve~(\ref{constraints}) in terms of $(Y,  {\widetilde Y})$ by setting:
\[
Y^1 = Y, \qquad Y^2 = - {\widetilde Y} - it^2, \qquad Y^3 = - n {\widetilde Y} - Y - it^1, \qquad Y^4 =   {\widetilde Y}.
\]
Integrating out the two massive $\Sigma$ f\/ields imposes two linear relations on the Fermi super\-f\/ields~$F^A$.\footnote{It is worth noting that the $\Sigma$ f\/ields are not always massive. By tuning the $E$-couplings appropriately, one can f\/ind new branches in which combinations of the~$\Sigma$ f\/ields becomes massless. The locus where such a branch meets a conventional Higgs branch usually corresponds to a bundle singularity with the rank of the bundle changing. There is a rich array of physical phenomena that can happen at these loci.} For simplicity, let us assume the gauge bundle is a deformation of the tangent bundle. In this case, we solve the constraints
\[
F^1 + F^3 + n F^4 =0, \qquad F^2 + F^4=0,
\]
in terms of
\[
F^1=F, \qquad F^2 = -{\widetilde F}, \qquad F^3 = -n{\widetilde F} - F, \qquad F^4 =  {\widetilde F},
\]
in parallel with the $Y^i$ discussion.
On the (2,2) locus, $\beta_{iA} = - {1\over \sqrt{2}} \delta_{iA}$. We will deform around this point. Expressed in terms of these f\/ields, the resulting superpotential takes the form
\begin{gather} \label{hirzepot}
W = {\mu\over \sqrt{2}} F \left[ e^{it^1 +Y + n {\widetilde Y}} -   e^{- Y} \right] + {\mu\over \sqrt{2}}  {\widetilde F} \left[ e^{it^2 + {\widetilde Y}}  + n e^{it^1 +Y + n {\widetilde Y}} -  e^{- {\widetilde Y}}  \right] + {\widetilde W},
\end{gather}
where deformations away from the (2,2) locus are captured in ${\widetilde W}$ with parameters ${\widetilde \alpha}$:
\begin{gather*}
{\widetilde W}  = {\mu\over \sqrt{2}} F \left[{\widetilde \alpha}_{11} e^{-Y}  +  {\widetilde \alpha}_{12} e^{it^1 +Y + n {\widetilde Y}} + {\widetilde \alpha}_{13} e^{- {\widetilde Y}} +  {\widetilde \alpha}_{14} e^{it^2 + {\widetilde Y}} \right] \\
\hphantom{{\widetilde W}  =}{}
+ {\mu\over \sqrt{2}}  {\widetilde F} \left[ {\widetilde \alpha}_{21} e^{ -Y}   +   {\widetilde \alpha}_{22} e^{it^1 +Y + n {\widetilde Y}} +{\widetilde \alpha}_{23} e^{ - {\widetilde Y}} +  {\widetilde \alpha}_{24} e^{ it^2+ {\widetilde Y}} \right].
\end{gather*}
If all ${\widetilde \alpha}=0$, we are on the (2,2) locus. It is important to stress that we could have considered models with no connection to the tangent bundle. As long as the rank of the chosen holomorphic bundle, given by the sequence~(\ref{geneuler}), is at least as large as the dimension of the target space, the structure of the dual theory is still determined by a superpotential of the form~(\ref{exact}). The form of the dual theory for models with ${\rm rk}(\cE) < {\rm rk}(T_V)$ will be described in Section~\ref{key}.

To obtain the quantum sheaf cohomology ring for this case, let us introduce ${\mathbb C}$-valued f\/ields:
\[
Z = e^{-Y}, \qquad {\widetilde Z} = e^{-\widetilde Y}.
\]
Solving the $F$ and $\widetilde F$ constraints of~(\ref{hirzepot}) in terms of these variables gives the ring relations. If all ${\widetilde \alpha}=0$, we f\/ind the quantum cohomology ring:
\[ %\label{hirzeqcoh}
Z^2 {\widetilde Z}^n = e^{it^1}, \qquad  {\widetilde Z}^{2} - n Z {\widetilde Z}  = e^{it^2}.
\]
As a basic check on this structure, we can see if this ring reduces to the classical cohomology ring of a Hirzebruch surface. The classical limit is obtained by taking the K\"ahler classes of both f\/iber and base to inf\/inity. This corresponds to the limit $e^{it^a} \rightarrow 0$ from which we see the classical cohomology ring of ${\mathbb F}_n$ emerge.

The more general quantum sheaf cohomology ring is given by solving the  $F$ and $\widetilde F$ constraints, including a non-vanishing $\widetilde{W}$ in~(\ref{hirzepot}). To make life simpler, let us set to zero the diagonal ${\widetilde \alpha}$ coef\/f\/icients that correspond to rescalings of the (2,2) couplings which already appear in~(\ref{hirzepot}):
\[
{\widetilde \alpha}_{11} = {\widetilde \alpha}_{12} =0, \qquad  {\widetilde \alpha}_{22} = {\widetilde \alpha}_{23}={\widetilde \alpha}_{24}=0.
\]
This leaves three deformation coef\/f\/icients. The quantum sheaf cohomology ring takes the form,
\[
Z^2 {\widetilde Z}^n - {\widetilde \alpha}_{13} Z {\widetilde Z}^{n+1} - {\widetilde \alpha}_{14} Z {\widetilde Z}^{n-1} e^{it^2} = e^{it^1},  \qquad
 {\widetilde Z}^{2} - n Z {\widetilde Z}  -  {\widetilde \alpha}_{21} Z  {\widetilde Z}  = e^{it^2}.
\]
This includes both a classical deformation of both relations which survives $e^{it^a} \rightarrow 0$, as well as a possible quantum deformation of the f\/irst relation proportional to $e^{it^2}$. These results should be contrasted with the analysis presented in Section~\ref{mathhirze}, noting that $q^a = e^{it^a}$.

\subsection{Key dif\/ferences between (0,2) and (2,2) models}\label{key}

There are signif\/icant dif\/ferences in the structure of the mirror depending on how the rank of the bundle $\cE$ compares with the rank of the tangent bundle of the target space $V$. These are uniquely (0,2) phenomena with no analogue in the (2,2) setting.

\subsubsection*{Models with $\boldsymbol{{\rm rk}(\cE) < {\rm rk}(T_V)}$}

In these cases, the dual theory is not a Landau--Ginzburg theory at all!  This can be seen from counting constraints. Imagine a dual model with $N$ $Y^i$ superf\/ields and $M$ $F^A$ Fermi superf\/ields and $r$ Abelian gauge-f\/ields. The
vacuum structure is determined by solving
the constraints
\[ \sum_{i=1}^N Q_i^a Y^i =-it^a, \qquad \sum_{A=1}^M Q_A^a F^A =0,
\]
with $N>M$. We are left with $N-r$ $Y$ variables, and $M-r$ Fermi
superf\/ields. A generic non-perturbative superpotential  of the
form $\mu \sum_{iA} \beta_{iA} F^A e^{-Y^i}$ imposes a further
$M-r$ constraints on the
$Y$ f\/ields. However, the potential now has f\/lat directions corresponding to the excitations of the massless
$N-M$ $Y$ f\/ields. The low-energy theory is not a Landau--Ginzburg theory with isolated vacua, but
a non-linear sigma model with the vacuum manifold as a target space.

On examining the sigma model metric on this space, we f\/ind a f\/inite distance singularity
at $Y=0$. This singularity signals a breakdown of worldsheet perturbation theory. On the singular locus, the
dilaton of the dual theory diverges. This same phenomena is not uncommon in dualities relating minimal models
and sigma models; for examples, see~\cite{Giveon:1994fu}. The physics of models in this class has yet to be deeply
explored.

\subsubsection*{Models with $\boldsymbol{{\rm rk}(\cE) \geq {\rm rk}(T_V)}$}

For the purpose of constructing realistic models of particle physics, these cases are the most interesting. The case of
 ${\rm rk}(\cE) = {\rm rk}(T_V)$ is the most heavily studied class of examples; particularly  the case of deformations of $T_V$. For models with   ${\rm rk}(\cE) > {\rm rk}(T_V)$, the dual theory is again a~Landau--Ginzburg theory with a superpotential given by~(\ref{exact}). However, counting constraints as above leads to the conclusion that there are generically no supersymmetric vacua, or only vacua at $Y^i \rightarrow \infty$ for all $i$. For models with all positive charges, there are generically no vacua at all. Supersymmetry is spontaneously broken in these models!

 Dynamical supersymmetry breaking is a phenomenon that can happen in (0,2) models even with simple target spaces like ${\mathbb P}^N$. In the original frame, the physics leading to supersymmetry breaking is non-perturbative and is typically seen in a large $N$ expansion. This is really just an illustration of the rich physics waiting to be understood in (0,2) f\/ield theories. In terms of f\/ield theory dynamics, (0,2) models are quite akin to ${\cal N}=1$ four-dimensional gauge theories, while (2,2) models are similar to ${\cal N}=2$ four-dimensional gauge theories. Supersymmetry breaking is just one example of this analogy. Of course, one can tune the $E$-couplings in the original model so that the dual theory has supersymmetric vacua. It would be very interesting to characterize the quite special bundles which actually preserve supersymmetry.

%\newpage
\section{Current and future directions}
\label{summary}

The class of (0,2) theories we described  of\/fer a rich set of structures of interest to both
mathematicians and physicists.
Mirror symmetry in the (0,2) arena is a part of a larger story which we expect will provide
new insights into the nature of quantum geometry.
 The study
of tangent bundle deformations of\/fers a tractable entry point into this
mysterious and tantalizing world, where we can usefully apply many lessons
from conventional (2,2) mirror symmetry.  Clearly the most important and
exciting direction is to generalize these studies to more general bundles,
but there are also important questions remaining even for
tangent bundle deformations.
We end by summarizing a few of those questions.
The list is by no means complete, but we hope it will whet the reader's
appetite for a deeper look at (0,2) theories.

\begin{enumerate}\itemsep=0pt
\item \textbf{Vanishing theorems for quantum obstructions.}
We discussed the issue of stability of (0,2) deformations,
citing the advantages of the GLSM deformations and results
of~\cite{Basu:2003bq,Beasley:2003fx, Silverstein:1995re}.  Using those
results, can we prove, say for the class of ref\/lexively plain models,
that (0,2) GLSM deformations are not quantum obstructed?

\item \textbf{Quantum sheaf cohomology for hypersurfaces and complete
intersections.}  Although quantum sheaf cohomology should exist for
much more general cases,  computations are currently well-understood
only for toric varieties.
For hypersurfaces in toric varieties, there is a
``quantum restriction'' formula~\cite{McOrist:2008ji}, but more work needs
to be done to test and generalize this proposal.

\item \textbf{Testing the conjectured (0,2) mirror map.}
It would be nice to develop techniques to test the (0,2) mirror map proposal
further.  A f\/irst step might be to check that other (non-principal)
components of the discriminant locus are also exchanged by the map.
A~more ambitious step would be to show the equality of A/2 and B/2
correlators.  This will require a (0,2) generalization of the theory of
toric residues, computations of quantum sheaf cohomology for hypersurfaces,
as well as properly understanding the operator map in the absence of
topological descent.

\item \textbf{Extending the (0,2) mirror map.}
The restriction to ref\/lexively plain polytopes, although perhaps elegant,
is clearly a computational crutch.  Can we develop a more general notion
of the map that applies to a wider class of GLSMs?  As suggested
in~\cite{Melnikov:2010sa}, a useful way to pursue this might be to study
``mirror subfamilies'' along the lines of~\cite{Morrison:1995yh}.

\item\textbf{``Quantum bundles'' and quantum geometry.}
A tantalizing feature of (0,2) theories, already apparent from the results on
singular loci, is that a sheaf corresponding to a~classically singular bundle
can nevertheless lead to a perfectly smooth conformal f\/ield theory.  Can we
develop a precise mathematical characterization of which sheaves have this
property?  What exactly are their mirrors?

\item\textbf{Deriving the (0,2) mirror map.}
Can the derivation of mirror pairs from worldsheet duality be extended to compact Calabi--Yau spaces? Any such extension  would provide a~proof of the mirror map for tangent bundle deformations, and hopefully,  provide a~ge\-ne\-ralization to a wider class of bundles.
\end{enumerate}

 \appendix
 \section{Superspace and superf\/ield conventions}
\label{superfieldconventions}

\subsection{Chiral and Fermi superf\/ields}

In this appendix, we summarize our notation and conventions used primarily in
Section~\ref{worldsheetduality}.
For a nice review of~(0,2) theories, see~\cite{McOrist:2010ae}.
Throughout this appendix, we will use the language of~(0,2) superspace
with coordinates $(x^+,x^-,\th^+,\thbar^+)$.
The worldsheet coordinates are def\/ined by $x^\pm = \hlf(x^0\pm x^1)$
so the corresponding derivatives $\del_\pm = \del_0 \pm \del_1$ satisfy
$\del_\pm x^\pm =1$.
We def\/ine the measure for Grassman integration so that
$\d^2\th^+ = \d\thbar^+ \d\th^+$ and  $ \int\d^2\th^+\, \th^+\thbar^+ = 1.$
The (0,2) super-derivatives
\begin{gather*}
D_+   =   \del_{\th^+}   -   i\thbar^+\del_+, \qquad
\Dbar_+   =   -\del_{\thbar^+}  +   i\th^+\del_+,
\end{gather*}
satisfy the usual anti-commutation relations
\begin{gather*}
\{D_+,D_+\}   =   \{\Dbar_+,\Dbar_+\}   =   0, \qquad
\{\Dbar_+,D_+\}   =   2i\del_+ .
\end{gather*}

In the absence of gauge f\/ields, (0,2) sigma models involve two sets of
superf\/ields: chiral superf\/ields annihilated by the $\Dbar_+$ operator,
\begin{gather*}
\Dbar_+\F^i   =   0,
\end{gather*}
and Fermi superf\/ields $\G^\a$ which satisfy,
\begin{gather*}
\Dbar_+\G^\a   =   \sqrt{2} E^\a,
\end{gather*}
where $E^\a$ is chiral: $\Dbar_+ E^\a=0$. These superf\/ields have the following
component expansions:
\begin{gather}
\F^i  =  \f^i   +   \sqrt{2}\th^+\j_+^i   -   i \th^+\thbar^+\del_+\f^i,
\label{chiral}\\
\G^\a  =  \g^\a   +   \sqrt{2}\th^+F^\a   -   \sqrt{2} \thbar^+ E^\a
-   i\th^+\thbar^+\del_+\g^\a.\nonumber
\end{gather}

If we omit superpotential couplings, the most general  Lorentz invariant (0,2) supersymmetric action involving only chiral and Fermi superf\/ields and their complex conjugates takes the form,
\begin{gather*} %\label{(0,2) sigma}
\L   =   -\hlf\int\d^2\th^+\left[{i\over2}K_i \del_- \F^i   -
{i\over2} K_{\ibar } \del_- {\bar \F}^{\ibar}   +
h_{\a\bar{\b}}\bar{\G}^{\bar \b}\G^\a   +
h_{\a\b}\G^\a\G^\b   +
h_{\bar{\a}\bar{\b}}\bar{\G}^{\bar{\a}}\bar{\G}^{\bar{\b}} \right].
\end{gather*}
The one-forms $K_i$ determine the metric;
the functions $h_{\a\b}$ and $h_{\a\bar{\b}}$ determine the bundle metric.

\subsection{Gauged linear sigma models}

We now introduce gauge f\/ields. For a general $U(1)^n$ Abelian gauge theory, we require a pair (0,2)  gauge superf\/ields $A^a$ and $V_-^a$ for each Abelian factor, $a=1,\ldots,n$. Let us restrict to $n=1$ for now. Under a super-gauge transformation, the vector superf\/ields transform as follows,
\begin{gather*}
\dd A  =  {i}(\bar{\La}   -   \La)/2, \qquad
\dd V_-  =  - \del_-(\La   +   \bar{\La})/2,
\end{gather*}
where the gauge parameter $\La$ is a chiral superf\/ield: $\Dbar_+ \La=0$.  In Wess--Zumino gauge, the gauge superf\/ields take the form
\begin{gather*}
A  =  \th^+\thbar^+ A_+, \qquad
V_-  =  A_-   -   2i\th^+\labar_-   -   2i\thbar^+\lambda_-   +
2\th^+\thbar^+ D,
\end{gather*}
where $A_\pm = A_0 \pm A_1$ are the components of the gauge f\/ield.
We will denote the gauge covariant derivatives by
\begin{gather*}
\cD_\pm   =   \del_\pm   +   i Q A_\pm
\end{gather*}
when acting on a f\/ield of charge $Q$. This allows us to replace our usual superderivati\-ves~$D_+$,~$\Dbar_+$ with gauge covariant ones
\begin{gather*}
\mathfrak{D}_+   =   \del_{\th^+}   -   i\thbar^+\cD_+, \qquad
\bar{\mathfrak{D}}_+   =   -\del_{\thbar^+}   +i   i\th^+\cD_+
\end{gather*}
which now satisfy the modif\/ied algebra
\begin{gather*}
\{\mathfrak{D}_+,\mathfrak{D}_+\}   =
\{\bar{\mathfrak{D}}_+,\bar{\mathfrak{D}}_+\}   =   0, \qquad
\{\bar{\mathfrak{D}}_+,\mathfrak{D}_+\}   =   2i\cD_+ .
\end{gather*}
We must also introduce the supersymmetric gauge covariant derivative,
\begin{gather*}
\nabla_-   =   \del_-   +   i Q V_-,
\end{gather*}
which contains $\cD_-$ as its lowest component. The gauge invariant Fermi
multiplet containing the f\/ield strength is def\/ined as follows,
\begin{gather*}
\Upsilon =[\bar{\mathfrak{D}}_+,\nabla_-]   =
\Dbar_+(\del_- A + i V_-)   =
-2\big(\lambda_-   -   i\th^+(D-iF_{01})   -   i\th^+\thbar^+\del_+\lambda_-\big).
\end{gather*}
Kinetic terms for the gauge f\/ield are given by
\begin{gather} \label{LU}
\L   =   {1\over8e^2}\int\d^2\th^+\, \bar{\Upsilon}\Upsilon
  =   {1\over e^2}\left(\hlf F_{01}^2   +
i\labar_-\del_+\lambda_-   +   \hlf D^2\right).
\end{gather}
Since we are considering Abelian gauge groups, we can also introduce an FI term with complex coef\/f\/icient $t=ir + {\th\over2\pi}$:
\begin{gather} \label{LFI}
{t\over4}\int\d\th^+ \Upsilon\Big|_{\thbar^+=0} + \text{c.c.} = -rD + {\th\over2\pi}F_{01}.
\end{gather}

In order to charge our chiral f\/ields under the gauge action,
we should ensure that they satisfy the covariant chiral constraint
$\mathfrak{\bar{D}}_+\Phi = 0$. Since
$\mathfrak{\bar{D}}_+ = e^{QA}\Dbar_+e^{-QA}$ it follows that $e^{QA}\Phi_0$
is a~chiral f\/ield of charge $Q$, where $\Phi_0$ is the neutral chiral f\/ield
appearing in~(\ref{chiral}). In components,
\[
\F  =  \phi  +  \sqrt{2} \theta^+ \j  -  i\th^+\thbar^+\cD_+\phi.
\]
The standard kinetic terms for charged chirals in (0,2) gauged linear sigma models (GLSMs) are
\begin{gather}
\L  =  {-i\over2}\int\d^2\th^+ \bar{\F}^i \nabla_- \F^i  \nonumber\\
\hphantom{\L}{} =  \left(-\big|\cD_\mu \phi^i\big|^2   +   \bar{\psi}_+i\cD_-\psi_+^i   -
\sqrt{2}iQ_i \bar{\phi}^i\lambda_-\j^i_+   +
\sqrt{2}iQ_i\phi^i\bar{\j}_+^i\labar_-   +   D Q_i \big|\phi^i\big|^2\right). \label{LPhi}
\end{gather}
Fermi superf\/ields are treated similarly. We promote them to charged f\/ields
by def\/ining $\G = e^{QA}\G_{0}$ so that in components
\begin{gather*}
\G   =   \g   +   \sqrt{2}\th^+F   +
\sqrt{2}\thbar^+E   -   i\th^+\thbar^+\cD_+\g,
\end{gather*}
where we have introduced a non-vanishing $E$ again.
If we make the standard assumption that~$E$ is a holomorphic function of the
$\F^i$ then the  kinetic terms for the Fermi f\/ields are:
\begin{gather}
\L  =  -\hlf\int\d^2\th^+\,  \bar{\G}^\a \G^\a
 =  \left(i\bar{\g}^\a\cD_+\g^\a   +   \big|F^\a\big|^2   -
\big|E^\a\big|^2   -   \bar{\g}^\a\del_i E^\a \j_+^i   -
\bar{\j}_+^i \del_\ibar \bar{E}^\a \g^\a\right).\label{LLa}
\end{gather}

\subsection{Superpotential couplings}

We can introduce superpotential couplings,
\begin{gather} \label{super}
S_J   =   -{1\over \sqrt{2}}\int\d^2x\d\th^+\, \G \cdot J(\F)   +
{\rm c.c.},
\end{gather}
supersymmetric if $E\cdot J=0$, which give a total bosonic potential
\[
V   =  \frac{D^2}{2 e^2} +   |E|^2   +   |J|^2.
\]
The action consisting of the terms~(\ref{LU}),~(\ref{LFI}),~(\ref{LPhi}),~(\ref{LLa})
and~(\ref{super}) comprises the standard (0,2) GLSM.

\subsection*{Acknowledgements}

S.S.~was supported in part by NSF Grant No.~PHY-0758029 and
NSF Grant No.~0529954.
E.S.~was supported in part by NSF grant PHY-1068725.  I.M.~and S.S.~would like to thank the Simons Center for Geometry and Physics for hospitality during the completion of this work.
\vspace{-2mm}

\pdfbookmark[1]{References}{ref}
\LastPageEnding

\end{document}